\documentclass[journal]{IEEEtran}

\usepackage{cite}

\usepackage[dvipsnames,x11names,table]{xcolor}
\usepackage{graphicx}
\usepackage{subcaption}
\usepackage{float}
\usepackage[font=small]{caption}

\usepackage{amsmath,amssymb,amsfonts}
\usepackage{breqn}

\usepackage{algorithmic}
\usepackage{array}
\newcolumntype{L}[1]{>{\raggedright\let\newline\\\arraybackslash\hspace{0pt}}m{#1}}
\newcolumntype{C}[1]{>{\centering\let\newline\\\arraybackslash\hspace{0pt}}m{#1}}
\newcolumntype{R}[1]{>{\raggedleft\let\newline\\\arraybackslash\hspace{0pt}}m{#1}}

\usepackage{url}
\usepackage{etoolbox}
\usepackage{hyperref}

\newif\ifdraftnotes
\draftnotestrue

\ifdraftnotes
  
  \newcommand{\meas}[1]{\textcolor{red}{\,[#1]}}
  \newcommand{\TODO}[1]{\textcolor{blue}{[TODO: #1]}}
\else
  
  \newcommand{\meas}[1]{}
  \newcommand{\TODO}[1]{}
\fi

\newcommand{\CB}{C_B}
\newcommand{\Vin}{V_{in}}
\newcommand{\VTRIP}{\Delta V_{TRIP}}
\newcommand{\tdelay}{t_{delay}}
\newcommand{\Ileak}{I_{leak}}
\newcommand{\IFE}{I_{FE}}
\newcommand{\Vth}{V_{TH}}
\newcommand{\Vleak}{V_{leaker}}
\begin{document}

\title{A FeFET Voltage-to-Time Converter with Offset Trim by Programmed Multilevel State in 28-nm CMOS}

\author{Jeries~Mattar,~\IEEEmembership{Student Member,~IEEE,}
        Hanaa Eqeiq,
        Stefan~D\"unkel,
        Halid~Mulaosmanovic,~\IEEEmembership{Senior Member,~IEEE,}
        Gunda Beernink,
        Sven~Beyer,~and~Nicolás~Wainstein,~\IEEEmembership{Member,~IEEE}
\thanks{J. Mattar and N. Wainstein are with the Andrew and Erna Viterbi Faculty of Electrical and Computer Engineering, Technion - Israel Institute of Technology, Haifa, Israel. Corresponding author: e-mail: (nicolasw@technion.ac.il).}
\thanks{H. Eqeiq was with the Andrew and Erna Viterbi Faculty of Electrical and Computer Engineering, Technion - Israel Institute of Technology, Haifa, Israel and is currently with Astera Labs, Israel.}
\thanks{S. D\"unkel, H. Mulaosmanovic, G. Beernink, and S. Beyer are with GlobalFoundries Fab1 LLC and Co. KG, Dresden, Germany.}
\thanks{This work was funded by the European Union within ``NextGeneration EU'', the Federal Ministry for Economic Affairs and Climate Action (BMWK) on the basis of a decision by the German Bundestag and by the State of Saxony with tax revenues based on the budget approved by the members of the Saxon State Parliament in the framework of “Important Project of Common European Interest - Microelectronics and Communication Technologies", under the project name “EUROFOUNDRY”, and by the Uzia Galil Memorial Fund. N. Wainstein is the Robert J. Shillman Fellow.
 }
}

\maketitle
\begin{abstract}
The offset and gain of voltage-to-time converters (VTCs) in time-domain circuits vary with process, supply, and temperature. This work presents a VTC whose nonvolatile input-referred offset trim is the programmed multilevel state of the ferroelectric field-effect transistor (FeFET) that converts the input. Fabricated in 28-nm CMOS, the 6.07-$\mu$m$^2$ VTC consists of a current-starved inverter with the FeFET and a parallel NMOS leaker in its tail, a capacitor bank, and a following inverter. The programmed state shifts the transfer curve onto the input range, leaving the gain to the capacitor bank and the leaker bias. Measured at 10~MS/s, centered states weaken the second harmonic from $-$12.5~dB in the low-threshold state to $-$32.2~dB and lower the static integral nonlinearity (INL) at 5~bit from 2.42 to 0.92~LSB. Post-layout simulations indicate that the proposed VTC can operate at 500~MS/s, drawing 2.0~$\mu$W from a 0.9-V supply, with nearly unchanged distortion and gain up to 700~MS/s. Across simulated process corners and $\pm$10\,\% supply variation, the input-referred offset is calibrated to within 10~mV and the gain to within 4\,\% of nominal by adjusting the programmed state and leaker bias.
\end{abstract}


\section{Introduction}

\IEEEPARstart{A}{s} CMOS technology scales into the nanometer regime, reduced supply voltages and degraded intrinsic transistor gain make high-precision voltage-domain analog signal processing increasingly difficult~\cite{annema2005,lewyn2009}. Circuits that represent signals in the time domain exploit the speed and area efficiency of deeply scaled logic~\cite{whitcombe2025} and have been demonstrated in high-speed data converters~\cite{liu2024,chen2023,wang2026,lee2026,wang2026tcas} and in accelerators that accumulate weighted delays for vector--matrix operations~\cite{mattar2026,wu2024,yang2021}. A time-domain analog-to-digital converter (ADC) typically consists of a voltage-to-time converter (VTC), whose transfer curve maps the input voltage to a delay or a pulse width, followed by a time-to-digital converter (TDC) that quantizes this interval. The performance of these ADCs is limited by the conversion gain, input range, and linearity of VTCs. Process, supply voltage, and temperature (PVT) shift the transfer curve of a VTC and change its gain; and thus, offset and gain trimming techniques are widely used~\cite{yi2021,ohhata2019,zhao2023,qchen2022,wang2026}. Beyond correction of these variations, the transfer curve must be centered on the input common-mode voltage and the gain adjusted to the input range after fabrication to accommodate different input conditions.

In these VTCs, the trim is stored outside the transistor by which the conversion delay is controlled. Load capacitors are selected digitally to trim the delay~\cite{yi2021} or gain~\cite{zhao2023}, the time offset is adjusted through a coded delay buffer~\cite{qchen2022}, and gain and offset are trimmed through bias voltages~\cite{ohhata2019}. A gain bias can also be adjusted by a background loop using digital-output statistics~\cite{wang2026}. Additional circuitry is required for each trim. In recent ADCs based on resistive random-access memory (ReRAM)~\cite{yang2025,hong2025adaptiveadc}, the transfer curve is stored across tens of ReRAM devices, in which individual ramp steps or quantization boundaries are encoded. These devices must therefore be reprogrammed to establish a new transfer curve.

Calibration of comparators and delay-element circuits through stored nonvolatile states has been demonstrated in prior work~\cite{wong2008comp,wong2009quant,nouri2021ring,mattar2026}. Since a static offset need not be updated between conversions, it is well suited to nonvolatile storage. In~\cite{wong2008comp}, comparator offset is canceled by charge stored on the floating input gates. In a flash quantizer, reference levels are stored as floating-gate (FG) charges to align the quantization range with the input signal~\cite{wong2009quant}. Ring-oscillator frequency is tuned through charge trapped in a high-$k$ gate dielectric~\cite{nouri2021ring}, while delay elements in a time-domain computing macro with binary inputs are trimmed through multilevel states (MLSs) of ferroelectric field-effect transistors (FeFETs)~\cite{mattar2026}. In these circuits, an individual decision level, frequency, or delay is set by each stored state. Placement of an analog-to-time transfer characteristic over its input range through the programmed state of the converting device has not been demonstrated yet.

FeFETs are nonvolatile memories with a ferroelectric layer of doped HfO$_2$ in the gate stack of a CMOS transistor~\cite{boscke2011,trentzsch2016}. The retained polarization stores the state, which is read as a threshold-voltage shift~\cite{mikolajick2020}. A gate pulse of lower amplitude or shorter duration than a full write pulse partially switches the layer, producing an MLS between the low- and high-threshold states~\cite{chowdhury2026}. Each state provides a distinct channel conductance at a fixed gate voltage, while the gate voltage controls the drain current. A FeFET can therefore serve as a memory, a tunable resistor, or a programmable voltage-controlled current source. Here, this current source is used in a VTC whose input drives the FeFET gate to control the discharge of a load capacitance. The programmed state shifts the transfer curve to center it on the input common-mode voltage, storing the offset trim in the converting transistor.

In~\cite{eqeiq2026}, we reported a variable-slope VTC in 28-nm CMOS with a FeFET tail in a current-starved inverter, demonstrating programmed dc transfer curves, capacitor-bank tuning, and a 4-bit linear response in the low-threshold-voltage (LVT) state. The programmed state was interpreted as gain control because the slope over a fixed input range decreased toward the high-threshold-voltage (HVT) state. This article shows that the state primarily provides input-referred offset trim, shifting the transfer curve at nearly constant gain. Most of the previously observed slope change results from the fixed input range approaching a flat region of the shifted curve. Gain is controlled by the capacitor bank and leaker bias.

Building on this interpretation, the article adds four contributions. (a) A compact model of the multilevel FeFET states and a closed-form transfer model of the VTC show that the state enters only the midpoint, while the bank and the leaker set the amplitude. (b) Measurements under a sine input in LVT and intermediate states show that the centered state weakens the second harmonic from $-$12.5~dB in LVT to $-$32.2~dB. The same measurements show that noise, not distortion, then limits the resolution. (c) Static measurements in six programmed states show that centering lowers the 5-bit integral nonlinearity (INL) from 2.42 to 0.92~least significant bit (LSB). (d) Post-layout simulations quantify the distortion versus state, input swing, and conversion rate, and show that reselecting the state together with the leaker bias recenters the transfer curve across process and supply. None of the model, the single-tone measurements, the per-state linearity, or the recentering across process and supply appeared in~\cite{eqeiq2026}.

The remainder of this article is organized as follows. Section~\ref{sec:fefet} reviews FeFET multilevel operation and presents the compact model and its calibration. The VTC, its transfer model, and programmed-state placement of the transfer curve are described in Section~\ref{sec:vtc}. Section~\ref{sec:sim} evaluates distortion versus state, input swing, and conversion rate through post-layout simulations, including recentering across process and supply. Measured static linearity, single-tone distortion, and gain adjustment are reported in Section~\ref{sec:results}. The results are discussed and compared with prior art in Section~\ref{sec:comparison}, followed by conclusions in Section~\ref{sec:conclusion}.

\section{Ferroelectric FETs, Multilevel States, and Compact Model}
\label{sec:fefet}
\subsection{Ferroelectric FETs}

A FeFET is a field-effect transistor whose gate stack contains a ferroelectric layer. At zero field, the layer retains a polarization between two fully switched values, and a field above its coercive field ($E_c$) switches it from one value to the other. The polarization is nonvolatile and sets the threshold voltage ($\Vth$) of the transistor, so the two fully switched values give the LVT and HVT states. A write pulse raises the field in the layer above $E_c$ and switches the polarization, whereas a read applies a gate voltage ($V_{G}$) that keeps the field below $E_c$ and senses the drain current ($I_{D}$) without switching the layer~\cite{mikolajick2020}. Ferroelectricity in doped HfO$_2$~\cite{boscke2011} made the layer compatible with CMOS, and FeFETs have since been embedded in a 28-nm high-k metal-gate process~\cite{trentzsch2016}.

Beyond the two fully switched states, the ferroelectric layer also holds intermediate polarizations. Its domains switch at different fields, so a gate pulse that switches only part of them leaves a polarization between the two, and the $\Vth$ takes an intermediate value~\cite{mulaosmanovic2015}. The switched fraction grows with the amplitude and the width of the pulse. Such MLSs have been demonstrated in FeFETs~\cite{mattar2026,chowdhury2026,muller2023percolation}.

\subsection{Compact Model}
\label{sec:fefet:model}

In this work, the multilevel $\Vth$ state of a single FeFET is the programmed variable that places the transfer curve of the VTC. A state is the polarization that the ferroelectric layer retains after a write, from HVT through the intermediate states to LVT. A model of these states must capture both the $\Vth$ shift set by the retained polarization and the intermediate states produced by partial switching.
Since the process design kit provides no FeFET model, a compact model is built by coupling a Preisach model of the ferroelectric capacitor~\cite{dragosits2000preisach,kuhn2000fecap} to the foundry n-channel transistor, and the parameters of the model are fitted to the measured MLSs. 

The ferroelectric layer forms a capacitor in series with the gate of the transistor, so the retained charge sets the potential of the internal node between them and with it the $\Vth$ of the transistor. The layer polarization, $P$, is given by
\begin{equation}
    P = k\,P_s \tanh\!\left[w\left(E \mp E_c\right)\right] + P_{off},
    \label{eq:preisach}
\end{equation}
as described in~\cite{miller1990fecap,dragosits2000preisach}. Here, $E$ is the effective field in the layer, $P_s$ and $P_r$ are the saturation and remanent polarization, and the upper sign holds on the ascending branch. The loop steepness ($w$) is $\ln[(P_s+P_r)/(P_s-P_r)]/(2E_c)$, while the scale factor ($k$) and the offset ($P_{off}$) are recomputed from a stack of turning points at every field reversal~\cite{miller1990fecap}. An excursion short of saturation then opens a minor branch and retains a polarization between the two remanent values, representing one intermediate state.

After the write pulse, the layer retains a polarization $P_0$ on its minor branch. Normalization between the HVT and LVT states defines the dimensionless state variable
\begin{equation}
    s = \frac{P_0 + P_r}{2P_r},
    \label{eq:state}
\end{equation}
where $s=0$ denotes HVT and $s=1$ denotes LVT. Through $k$ and $P_{off}$, $s$ selects the minor branch of \eqref{eq:preisach}, determining the $I_{D}$--$V_{G}$ characteristic from the retained polarization. This polarization adds a fixed charge to the internal node between the layer and the transistor. With state-independent layer and gate capacitances assumed in the model, programming shifts the characteristic along the $V_{G}$ axis without changing its shape. Voltage division across the layer also makes the characteristic rise more gradually than that of the bare transistor. Polarization changes only when the field crosses $E_c$ in the model, leaving the programmed state unchanged during reads below $E_c$.

\subsection{Calibration}
\label{sec:fefet:calibration}

\begin{figure}[!t]
    \centering
    \includegraphics[width=0.9\columnwidth,trim={1cm 0.6cm 0.6cm 0.6cm},clip]{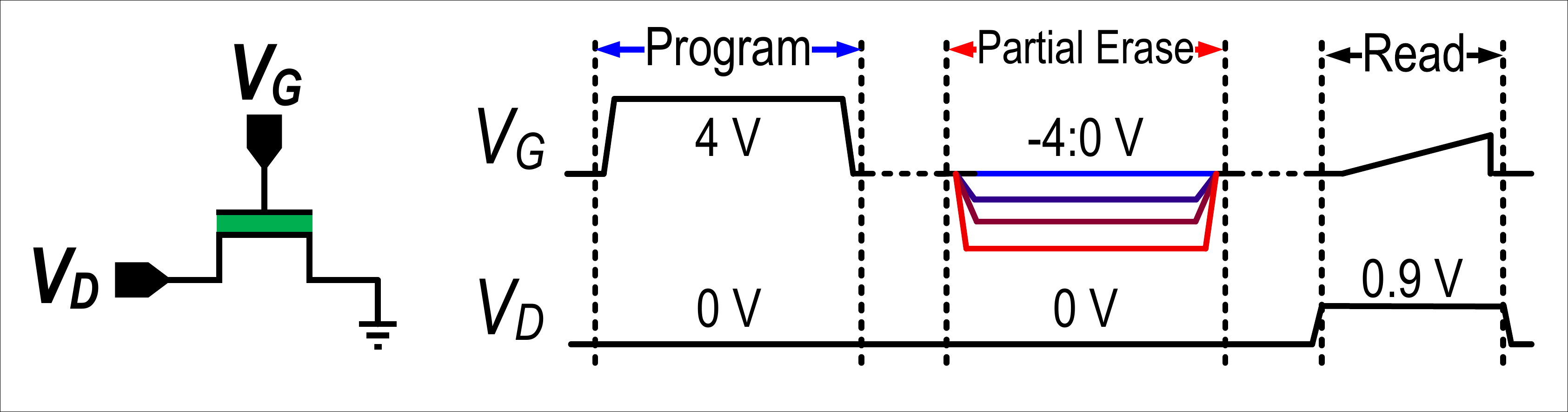}
    \caption{Waveform diagram illustrating the sequence of LVT programming, partial erase to an MLS, and read.}
    \label{fig:fefet_setup}
\end{figure}

\begin{figure}[!t]
    \centering
    \includegraphics[width=0.75\columnwidth,trim={0.9cm 0.9cm 0.9cm 0.9cm},clip]{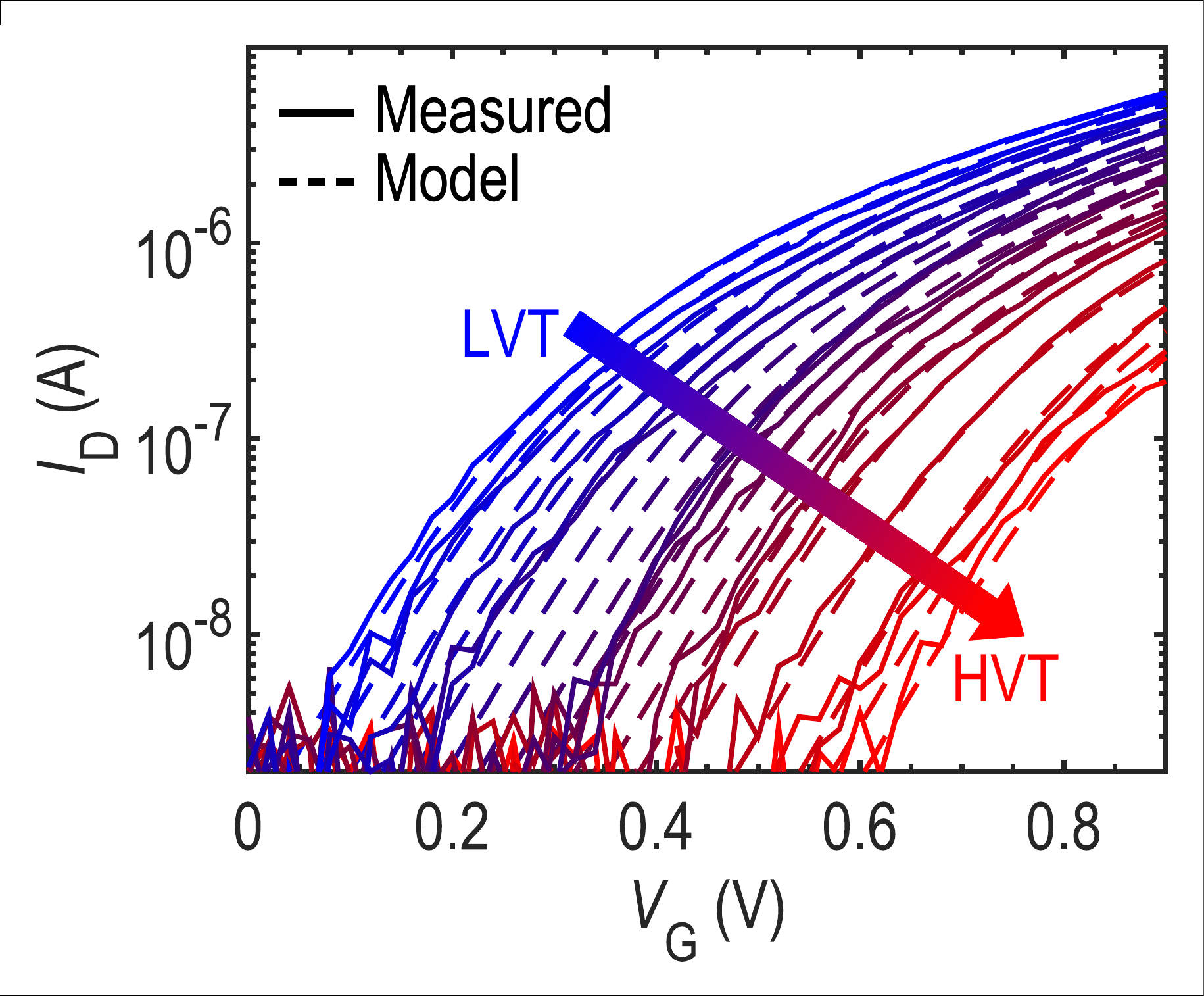}
    \caption{Measured (solid) and simulated (dashed) $I_{D}$ against $V_{G}$ with $V_{D}$ at 0.9~V for MLSs from LVT toward HVT.}
    \label{fig:fefet_model}
\end{figure}

The model is calibrated against MLSs measured on an n-channel FeFET in GlobalFoundries 28SLPe~\cite{trentzsch2016} with a gate width and length of 450~nm. The write and read sequence is shown in Fig.~\ref{fig:fefet_setup}. A gate pulse of $+$4~V first programs the FeFET to LVT, and a negative gate pulse then erases it partially toward HVT, with the drain voltage ($V_{D}$) at 0~V during both pulses. The amplitude of the erase pulse is stepped between 0 and $-$4~V, which leaves MLSs between LVT and HVT. Each state is then read by sweeping $V_{G}$ from 0 to 0.9~V with $V_{D}$ at 0.9~V, which gives its $I_{D}$--$V_{G}$ characteristic.

The parameters of the model are fitted to the measured MLSs, with a single parameter set shared by all states and only the state variable $s$ of \eqref{eq:state} changing between them. Fig.~\ref{fig:fefet_model} overlays the simulated characteristics for a sweep of $s$ from HVT to LVT on the measured MLSs. The simulated states follow the measured ones over the upper decade of current, and the states differ mainly by a shift in $\Vth$, as the retained polarization predicts. The model thus reproduces the multilevel behavior of the FeFET in simulation, and $s$ becomes the variable that programs the FeFET.

\section{Reconfigurable Voltage-to-Time Converter}
\label{sec:vtc}

\begin{figure}[!t]
    \centering
    \includegraphics[width=0.9\columnwidth,trim={0.6cm 0.6cm 0.6cm 0.6cm},clip]{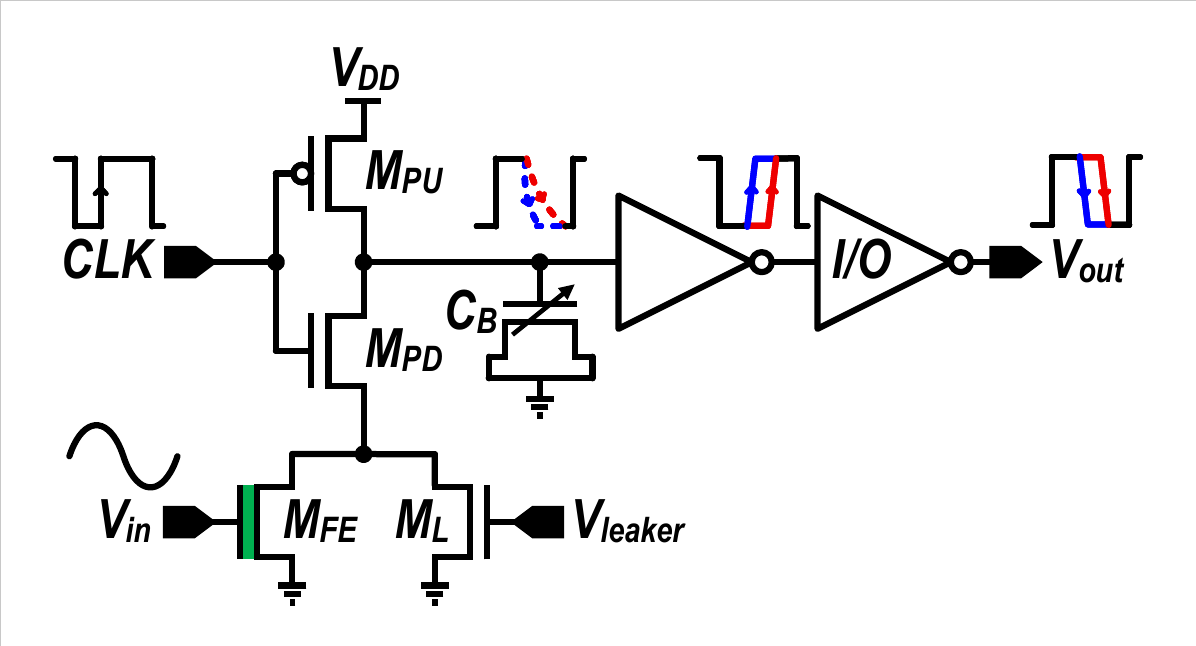}
    \caption{Schematic of the proposed FeFET-based VTC.}
    \label{fig:arch}
\end{figure}

The schematic of the proposed VTC is shown in Fig.~\ref{fig:arch}. The VTC consists of a clocked current-starved inverter, a 3-bit capacitor bank at its output, and a following inverter whose output an I/O driver carries off chip. A FeFET ($M_{FE}$) in parallel with an NMOS leaker ($M_{L}$) forms the tail of the current-starved inverter. The analog input ($\Vin$) drives the FeFET gate, and the leaker gate is held at a dc bias ($V_{leaker}$). The programmed MLS of the FeFET shifts the transfer curve, while the capacitor bank and the leaker bias set its gain. In this section, the parameters of a transfer curve are first presented, and the proposed VTC circuit, its transfer model, and the role of the programmed state are then described.

\subsection{VTC Types and Transfer-Curve Parameters}
\label{sec:vtc:fundamentals}

VTCs can be categorized as variable-slope or constant-slope converters, depending on whether the input sets the slope or the initial voltage of the ramp on a load capacitance. In a variable-slope VTC~\cite{yi2021,ohhata2019}, the input drives the gate of a current-source transistor and sets the current that charges or discharges the load capacitance, so the delay is inversely proportional to that current. This current is exponential in the gate voltage below threshold and grows more slowly than exponentially above it, and the delay is nonlinear in both regions. In a constant-slope VTC~\cite{qchen2022,jssc2023gro}, the sampled input sets the initial voltage on the load capacitance. A constant current then ramps that voltage to a fixed threshold, so the delay is linear in the input. Hence, the variable-slope VTC needs no sampling switch, whereas the constant-slope VTC is more linear~\cite{qchen2022}. In this work, we use a FeFET as the current-source transistor implementing a variable-slope VTC.

For either VTC type, the transfer curve is described by the midpoint and conversion gain. The midpoint is the input at which the output time crosses the middle of its swing. Its design value is the input common-mode voltage, so that the curve is centered on the input range. A shift of the midpoint from its design value is an input-referred offset. The conversion gain is the slope of the output time against the input, and a deviation of the gain from its design value is a gain error. In a variable-slope VTC, this slope changes along the curve, so the gain over an input range is taken either as the output swing divided by the input swing or as the slope of a least-squares line over that range.

A shift of the $\Vth$ of the current-source transistor displaces the midpoint of a variable-slope VTC and causes an input-referred offset. A shift of the switching threshold of the following stage changes the voltage swing of each conversion, scales every delay, and causes a gain error. Since both thresholds vary with PVT, variable-slope VTCs presented in~\cite{yi2021,ohhata2019,zhao2023,wang2026} trim their gain and their offset. In~\cite{yi2021,wang2026} and in the constant-slope VTC of~\cite{qchen2022}, the offset is trimmed as a time offset. Such a trim adds the same delay to every output and leaves the midpoint in place.

\subsection{Circuit Implementation}
\label{sec:vtc:cell}

Our VTC, shown in Fig.~\ref{fig:arch}, is implemented in GlobalFoundries 28SLPe, using an n-channel FeFET with gate width and length of 450~nm. The FeFET and the leaker both connect the source of the clocked pull-down transistor to ground, so the discharge current is the sum of their drain currents. With no sampling switch before the FeFET, the input drives its gate on each conversion. The I/O driver is used only for measurement and is excluded from the area and power of the VTC.

The clock drives the gates of the pull-up ($M_{PU}$) and pull-down ($M_{PD}$) transistors, and each clock period gives one conversion. In the precharge phase, $M_{PU}$ charges the output node of the current-starved inverter to the supply. In the discharge phase that follows, the tail discharges that node through $M_{PD}$. The conversion delay ($\tdelay$) is the time this discharge takes to reach the switching threshold of the following inverter, which lies a voltage swing ($\VTRIP$) below the supply:
\begin{equation}
    \tdelay(\Vin)=\frac{\CB\,\VTRIP}{\Ileak+\IFE(\Vin)}.
    \label{eq:tdelay}
\end{equation}
Here, $\CB$ is the capacitance of the capacitor bank together with the parasitic capacitance of the output node, $\Ileak$ is the leaker current, and $\IFE(\Vin)$ is the FeFET drain current.

The low pulse at the output of the I/O driver ($V_{out}$) starts when the following inverter switches and ends with the discharge phase, so only its falling edge moves with the input. Its width ($t_{PW}$) is $\tdelay$ subtracted from a fixed offset ($K$) set by the length of the discharge phase:
\begin{equation}
    t_{PW}(\Vin)=K-\tdelay(\Vin).
    \label{eq:width}
\end{equation}
A larger $\IFE$ shortens the delay and widens the pulse, so $t_{PW}$ rises with $\Vin$. A conversion completes only if the output of the current-starved inverter reaches the switching threshold within the discharge phase.

\subsection{Transfer Model}
\label{sec:transfer}

In weak inversion, the FeFET drain current is $\IFE=I_S\exp[(\Vin-\Vth)/(nU_{T})]$, where $\Vth$ is the programmed threshold voltage, $I_S$ a characteristic current proportional to $W/L$, $n$ the subthreshold slope factor, and $U_{T}$ the thermal voltage. With this current, \eqref{eq:tdelay} and \eqref{eq:width} give the transfer curve
\begin{equation}
    t_{PW}(\Vin) = K - \frac{B}{1+\exp\!\left[(\Vin-V_{0})/(nU_{T})\right]}~,
    \label{eq:logistic}
\end{equation}
whose amplitude is $B=\CB\VTRIP/\Ileak$ and whose midpoint is
\begin{equation}
    V_{0} = \Vth + nU_{T}\ln\!\left(\Ileak/I_{S}\right).
    \label{eq:v0}
\end{equation}
At this midpoint, $\IFE$ equals $\Ileak$ and the conversion delay is $B/2$. At low inputs, the leaker alone carries the discharge current, so the delay approaches $B$ and the transfer curve flattens toward $K-B$. At high inputs, $\IFE$ far exceeds $\Ileak$, so the delay approaches zero and the curve flattens toward $K$. The curve is odd-symmetric about its midpoint, where the conversion gain peaks at $B/(4nU_{T})$.

The model of \eqref{eq:logistic} holds only while the FeFET stays in weak inversion. In a low-threshold state, the FeFET leaves weak inversion as the input rises toward the supply. At the highest inputs, it also enters the triode region during the discharge phase, since its drain cannot rise above the clock high level minus the $\Vth$ of the pull-down transistor. Both effects flatten the upper part of the transfer curve and move its midpoint away from $V_{0}$. Since these effects act mainly at the upper end of the curve, the model still gives its overall shape and shows how the programmed threshold, the leaker current, and the capacitance $\CB$ set its midpoint and amplitude.

\subsection{Programmed State and Transfer-Curve Placement}
\label{sec:vtc:roles}
\label{sec:vtc:controls}

In the compact model of Section~\ref{sec:fefet}, the retained polarization sets $\Vth$ by displacing the whole drain-current characteristic of the FeFET. The divider that the ferroelectric layer forms with the gate capacitance gives this characteristic the same shape in every state. A change of state thus shifts $V_{0}$ of \eqref{eq:v0} and every point of the transfer curve by the same step as $\Vth$, while $B$ contains no FeFET term. The programmed state acts as an input-referred offset trim at nearly constant gain. The bank code and the leaker bias set the gain. The capacitor bank sets $\CB$, which scales $B$ and the gain but does not enter \eqref{eq:v0}. The leaker bias sets $\Ileak$, which lowers $B$ and the gain and raises $V_{0}$ of \eqref{eq:v0} as the bias rises.

For a given leaker bias, the programmed state places the midpoint of the transfer curve relative to a fixed input range. A state that centers the midpoint on that range aligns the steepest part of the curve with the center of the range, while a state whose midpoint is far from the center puts part of the range on a flat end. This placement controls the distortion and the linearity of the conversion.

\section{Distortion Control and Transfer Recentering by Programmed State}
\label{sec:sim}

To quantify how the programmed MLS places the transfer curve and controls its distortion, post-layout simulations of the VTC use the compact model of Section~\ref{sec:fefet}. Unless otherwise stated, every simulation is at the design point, the typical corner (TT) at 0.90~V and 70~$^\circ$C. The leaker bias is held at 0.58~V and the capacitor bank at code 0, its smallest capacitance. The input is 0.40~V peak-to-peak centered on 0.60~V, and the clock runs at 500~MHz with a 1.6~ns discharge phase in the 2~ns period. With ideal clock and input sources, the VTC draws 2.0~$\mu$W from the 0.90~V supply, 4.0~fJ per conversion. The distortion of the width $t_{PW}$ of \eqref{eq:width} is reported as its spurious-free dynamic range (SFDR) and total harmonic distortion (THD) for a sine at 5/128 of the conversion rate, 19.53~MHz at the design point. In this section, the programmed states and the resolution, the input swing and the conversion rate, and the recentering across process and supply are presented.

\subsection{Programmed States and Resolution}
\label{sec:sim:state}

\begin{figure}[!t]
    \centering
    \includegraphics[width=0.75\columnwidth, trim={1.6cm 1.4cm 1.4cm 1.4cm}, clip]{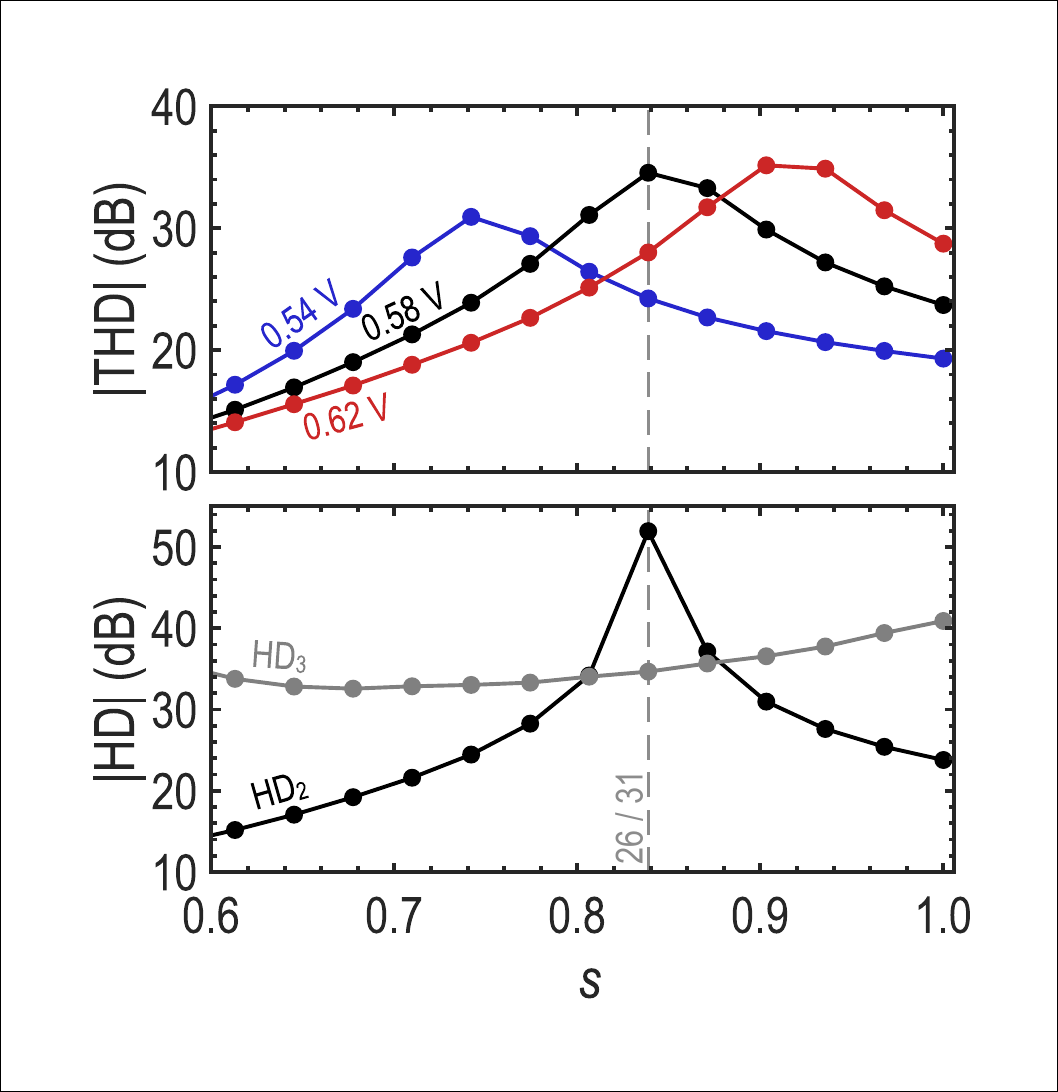}
    \caption{Simulated $|\text{THD}|$ at leaker biases of 0.54, 0.58, and 0.62~V (top) and second and third harmonics, $|\mathrm{HD}_2|$ and $|\mathrm{HD}_3|$, at 0.58~V (bottom) versus the programmed state $s$. The vertical dashed line marks $s=26/31$, the optimum at 0.58~V.}
    \label{fig:states}
\end{figure}

In the model, the state variable $s$ of \eqref{eq:state} is continuous between HVT and LVT, while a fabricated FeFET holds a limited number of distinct states. Single FeFETs have been programmed to 32 or more such states~\cite{jerry2017,chowdhury2026}. In multilevel FeFET memory arrays, device-to-device variation limits the number of states stored in each device~\cite{muller2023percolation}. The VTC instead needs only the programmed state that centers its transfer curve on the input swing. To find that state, the simulations sweep $s$ from HVT to LVT through 32 values in steps of $1/31$ at $\Vleak$ of 0.54, 0.58, and 0.62~V (see Fig.~\ref{fig:states}). With fewer values, the programmed state still places the transfer curve, controls its distortion, and recenters it. The number of values sets only the shift of the midpoint per step of $s$, and with it the midpoint offset that remains after the recentering of Section~\ref{sec:sim:recenter}.

At each $\Vleak$, one state minimizes the THD, and this optimum moves toward LVT as the bias rises, from $s=23/31$ at 0.54~V to $28/31$ at 0.62~V. At 0.58~V, the optimum $s=26/31$ improves the simulated THD by 10.9~dB with the bias unchanged, from $-$23.7~dB at LVT to $-$34.6~dB. From LVT to this optimum, the second harmonic weakens from $-$23.8 to $-$52.0~dB while the third grows from $-$40.9 to $-$34.7~dB (see Fig.~\ref{fig:states}). At the optimum, the third harmonic is the largest and sets both the THD and the SFDR.

The two harmonics follow from the shape of the transfer curve of \eqref{eq:logistic}, which flattens toward both ends and is odd-symmetric about its midpoint. An input above the midpoint raises the width exactly as much as an input equally far below lowers it. A sine centered on the midpoint is therefore compressed equally at its two peaks, which cancels the second harmonic and leaves the third. Off the midpoint, one peak is compressed more than the other, producing the second harmonic that sets the THD at LVT. At all three biases, the optimum is the state of least second harmonic (see Fig.~\ref{fig:states} at 0.58~V), and this state places the midpoint nearest the center of the input swing.

In the transfer model of Section~\ref{sec:transfer}, the midpoint $V_0$ of \eqref{eq:v0} is the input at which the FeFET current equals the leaker current. A higher leaker bias raises the leaker current and with it the midpoint, and a lower programmed threshold returns the midpoint to the center, explaining the shift of the optimum toward LVT. The higher leaker current also lowers the amplitude of \eqref{eq:logistic} and with it the gain, taken here as the swing of the width divided by the swing of the input. At 0.62~V instead of 0.58~V, this gain at the optimum falls by 43\,\% from 0.60 to 0.34~ns/V while the THD improves by 0.6~dB. The bias is held at 0.58~V to preserve the gain, and the programmed state places the midpoint of the transfer curve for that bias.

The resolution of the VTC is expressed by its effective number of bits (ENOB), set by the signal-to-noise-and-distortion ratio (SNDR), which combines the THD with the signal-to-noise ratio (SNR) of $t_{PW}$,
\begin{align}
    \mathrm{ENOB} &= \frac{\mathrm{SNDR}-1.76}{6.02}, \label{eq:enob}\\
    \mathrm{SNDR} &= -10\log_{10}\!\left(10^{\mathrm{THD}/10}+10^{-\mathrm{SNR}/10}\right), \label{eq:sndr}\\
    \mathrm{SNR} &= 20\log_{10}\frac{\Delta t_{PW}}{2\sqrt{2}\,\sigma_{PW}}. \label{eq:snr}
\end{align}
The THD, SNDR, and SNR are in dB and the ENOB in bits. Here, $\Delta t_{PW}$ is the peak-to-peak swing of $t_{PW}$ under the input sine, so the rms of its sine component is $\Delta t_{PW}/(2\sqrt{2})$. The rms noise $\sigma_{PW}$ is the standard deviation of $t_{PW}$ over repeated conversions of a constant input at the center of the swing. Transient-noise simulations with ideal clock and input sources give a $\sigma_{PW}$ of 4.5~ps, which stays within 5\,\% of that value as the bandwidth of the noise sources is raised from 5 to 500~GHz. With a $\Delta t_{PW}$ of 0.241~ns, the SNR is 25.5~dB, and with the distortion added, the SNDR is 25.0~dB and the ENOB 3.9~bit. Since the distortion alone gives 5.45~bit, the noise sets the resolution.

\subsection{Distortion Versus Input Swing and Conversion Rate}
\label{sec:sim:distortion}

\begin{figure}[!t]
    \centering
    \begin{subfigure}{0.85\columnwidth}
        \centering
        \includegraphics[width=0.9\columnwidth, trim={1.6cm 1.5cm 1.0cm 1.5cm}, clip]{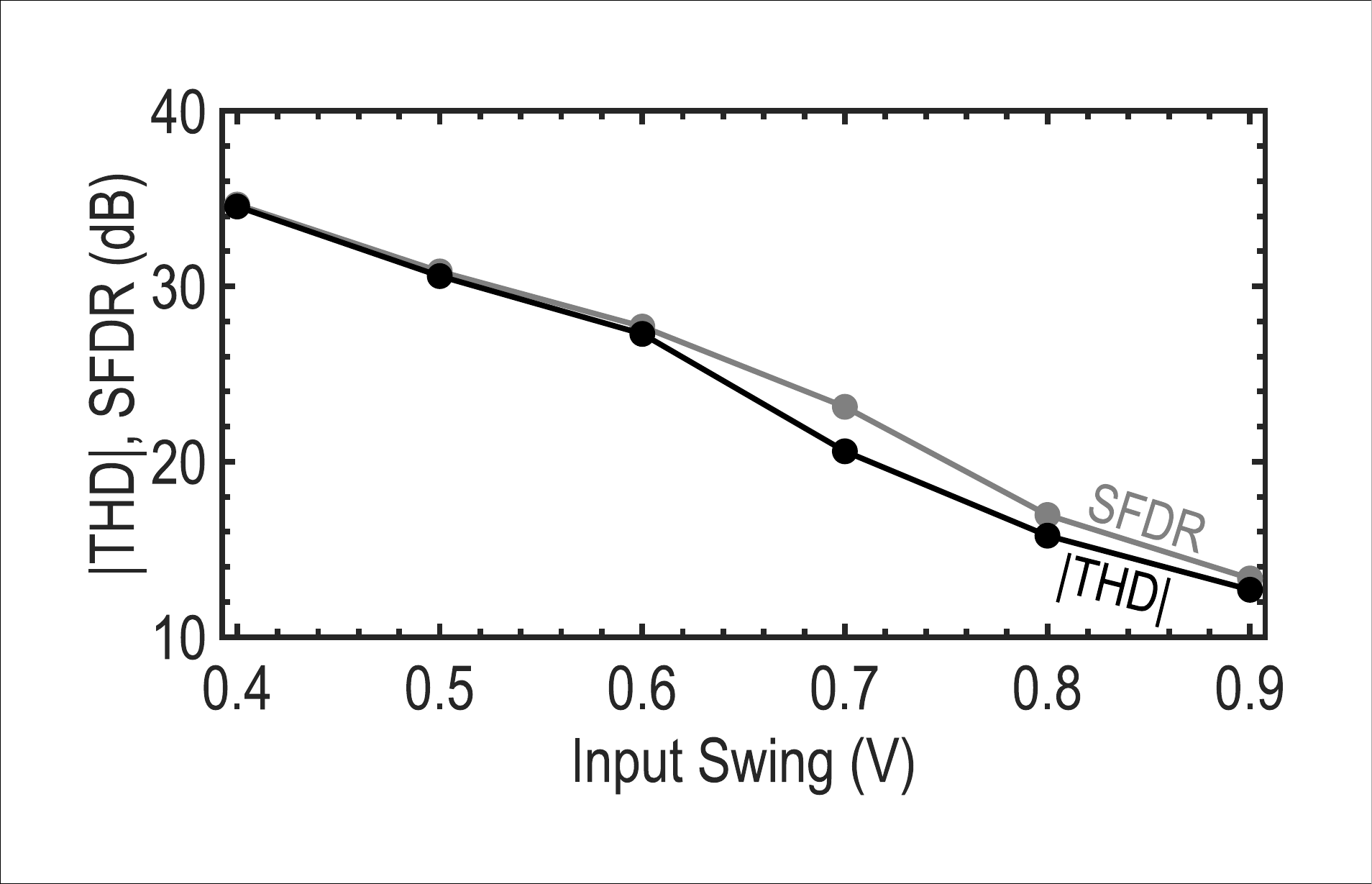}
        \vspace*{-0.5em}
        \caption{}
        \label{fig:sim:swing}
    \end{subfigure}\\[0pt]
    \begin{subfigure}{0.85\columnwidth}
        \centering
        \includegraphics[width=0.9\columnwidth, trim={1.5cm 1.5cm 1.5cm 1.5cm}, clip]{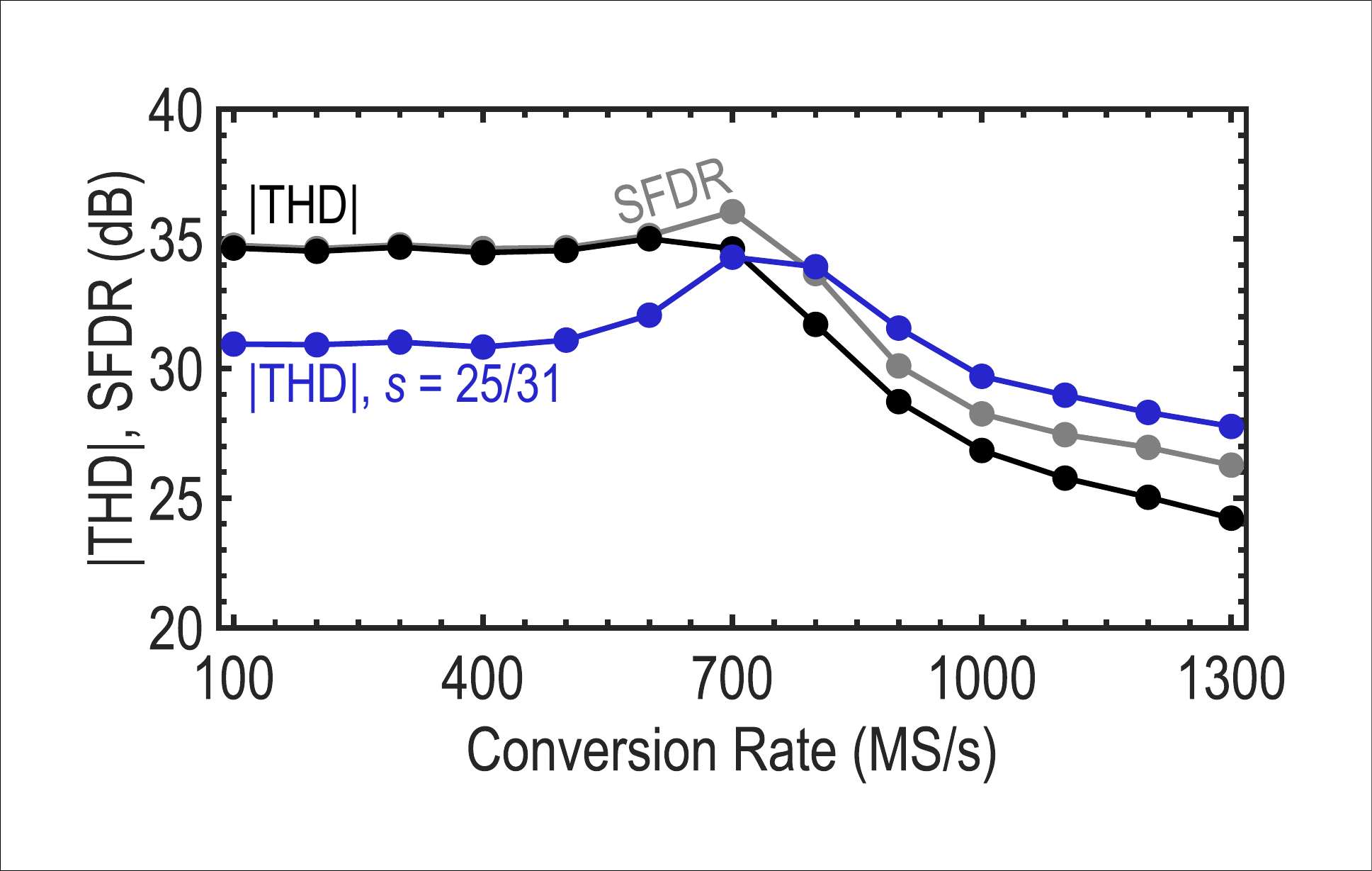}
        \vspace*{-0.5em}
        \caption{}
        \label{fig:sim:rate}
    \end{subfigure}
    \caption{Simulated $|$THD$|$ and SFDR at $s=26/31$ versus (a) peak-to-peak input swing and (b) conversion rate, with $|$THD$|$ at $s=25/31$, the state of least distortion above 700~MS/s.}
    \label{fig:sim}
\end{figure}

With the state at $s=26/31$, the peak-to-peak input swing is stepped from 0.40 to 0.90~V, each swing at its center of least distortion [see Fig.~\ref{fig:sim}(a)]. That center is 0.60~V up to a 0.60~V swing, and wider swings are centered lower to keep the input within the 0.90~V supply. The simulated THD stays below $-$30~dB up to a 0.50~V swing and degrades above that swing as the wider sine reaches the flat ends of the transfer curve. A shift of the center of the 0.40~V swing by 50~mV from 0.60~V degrades the THD by 3.5 to 6.1~dB, and a shift of the midpoint by 50~mV against a fixed input swing degrades it by the same amount.

The conversion rate is stepped from 100 to 1300~MS/s with the discharge phase held at 80\,\% of the period [see Fig.~\ref{fig:sim}(b)]. Up to 700~MS/s, the simulated THD varies by less than 0.6~dB and the gain by less than 3\,\%. Above that rate, the transfer curve shifts and steepens as the rate rises, the second harmonic returns, and the THD degrades to $-$24.2~dB at 1300~MS/s, where every conversion still completes. A state one step toward HVT ($s=25/31$) recenters the transfer curve, improving the THD by 2.2 to 3.5~dB from 800 to 1300~MS/s while changing the gain by less than 2\,\%.

\subsection{Recentering Across Process and Supply}
\label{sec:sim:recenter}

\begin{figure*}[!t]
    \centering
    \begin{subfigure}{0.32\textwidth}
        \centering
        \includegraphics[width=\linewidth]{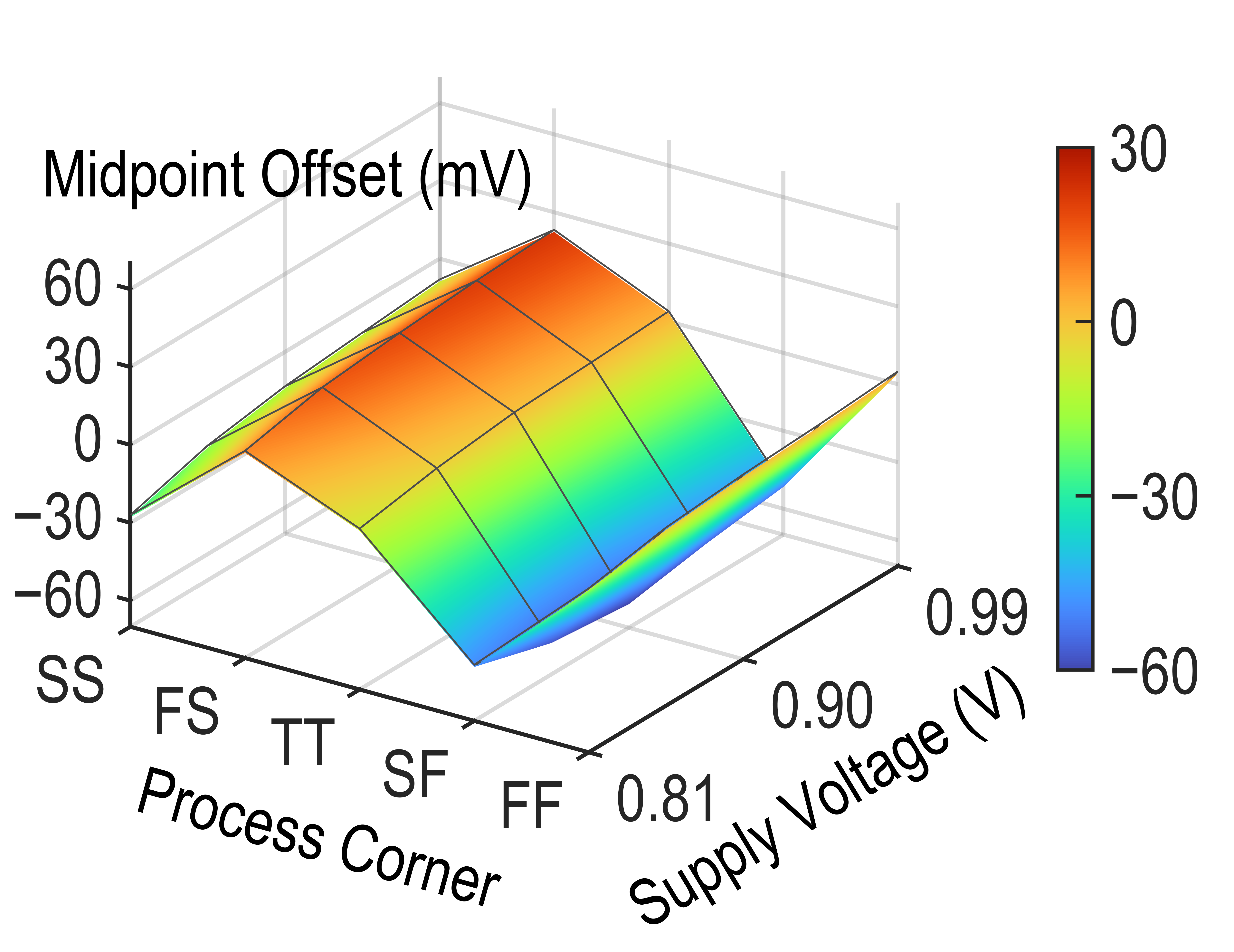}
         \vspace*{-1.8em}
        \caption{}
        \label{fig:pvt:a}
    \end{subfigure}\hfill
    \begin{subfigure}{0.32\textwidth}
        \centering
        \includegraphics[width=\linewidth]{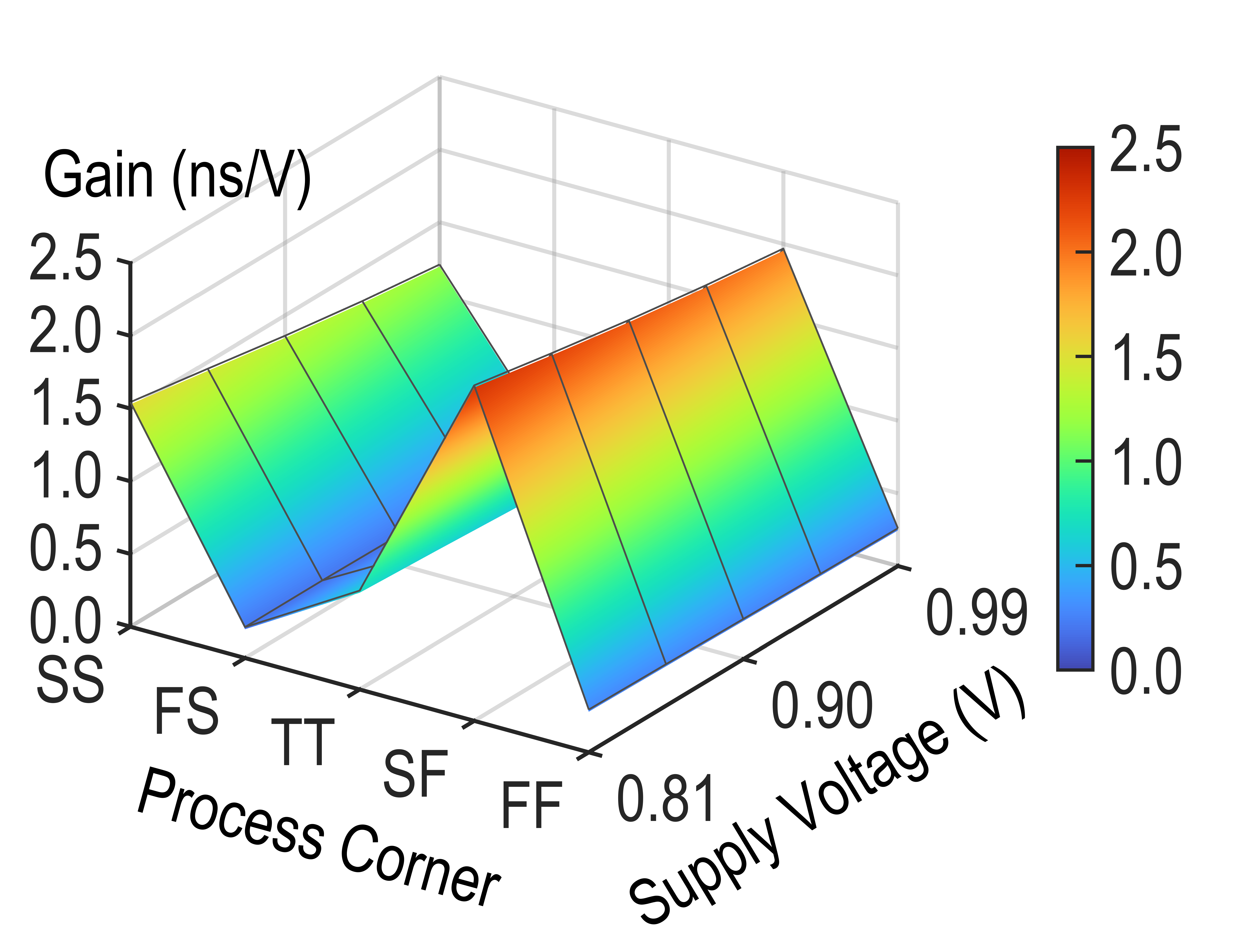}
         \vspace*{-1.8em}
        \caption{}
        \label{fig:pvt:b}
    \end{subfigure}\hfill
    \begin{subfigure}{0.32\textwidth}
        \centering
        \includegraphics[width=\linewidth]{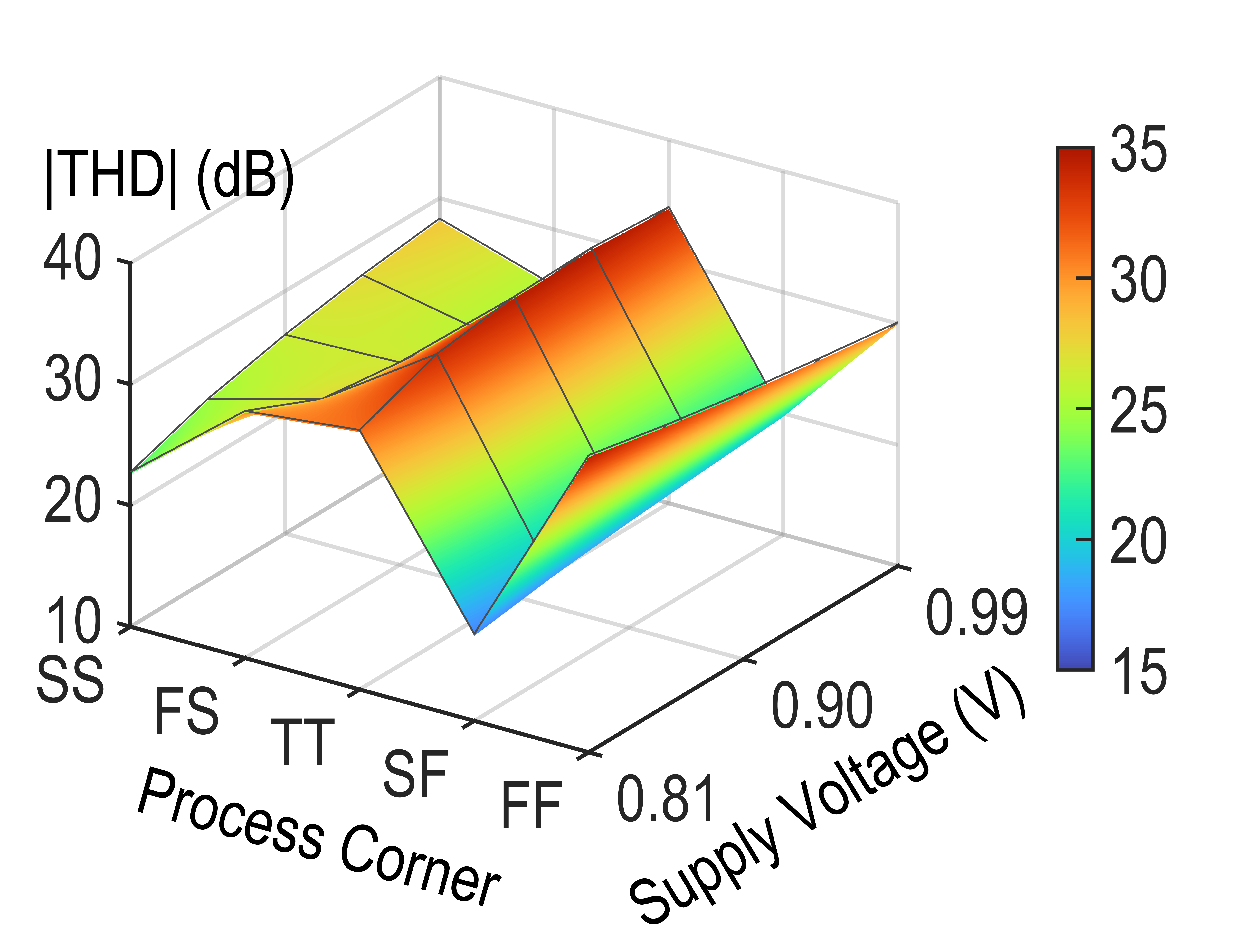}
         \vspace*{-1.8em}
        \caption{}
        \label{fig:pvt:c}
    \end{subfigure}\\[0pt]
    \begin{subfigure}{0.32\textwidth}
        \centering
        \includegraphics[width=\linewidth]{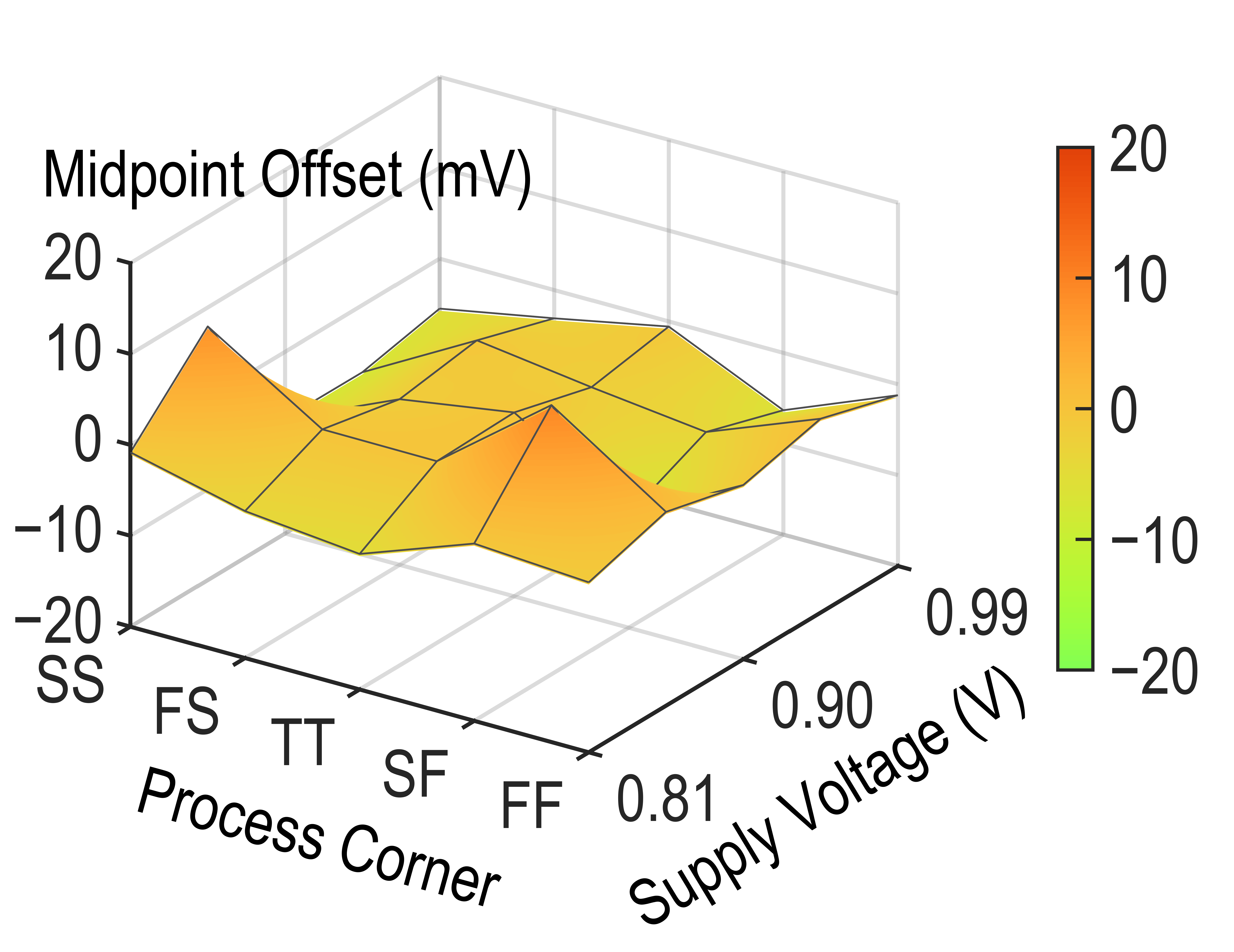}
        \vspace*{-1.8em}
        \caption{}
        \label{fig:pvt:d}
    \end{subfigure}\hfill
    \begin{subfigure}{0.32\textwidth}
        \centering
        \includegraphics[width=\linewidth]{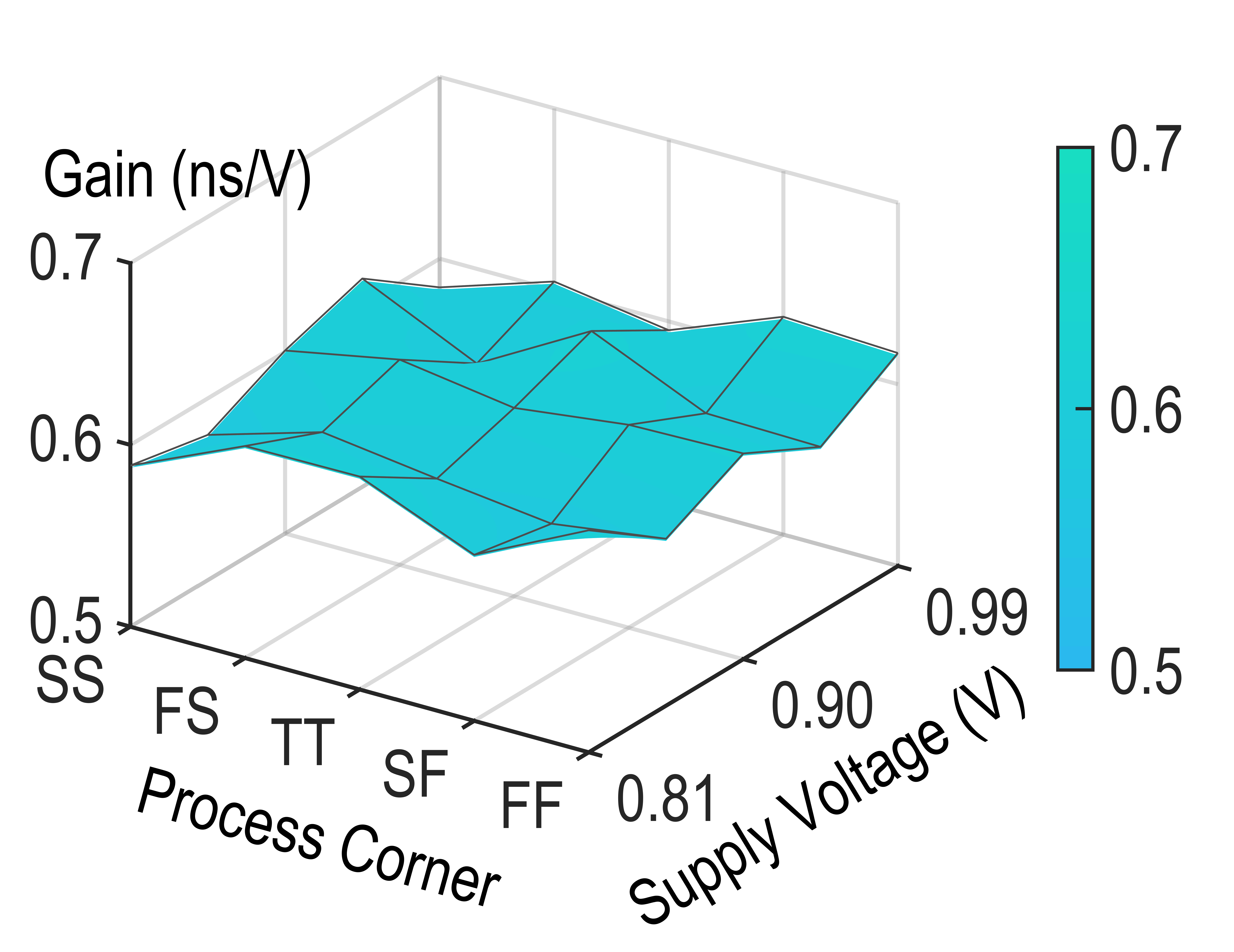}
        \vspace*{-1.8em}
        \caption{}
        \label{fig:pvt:e}
    \end{subfigure}\hfill
    \begin{subfigure}{0.32\textwidth}
        \centering
        \includegraphics[width=\linewidth]{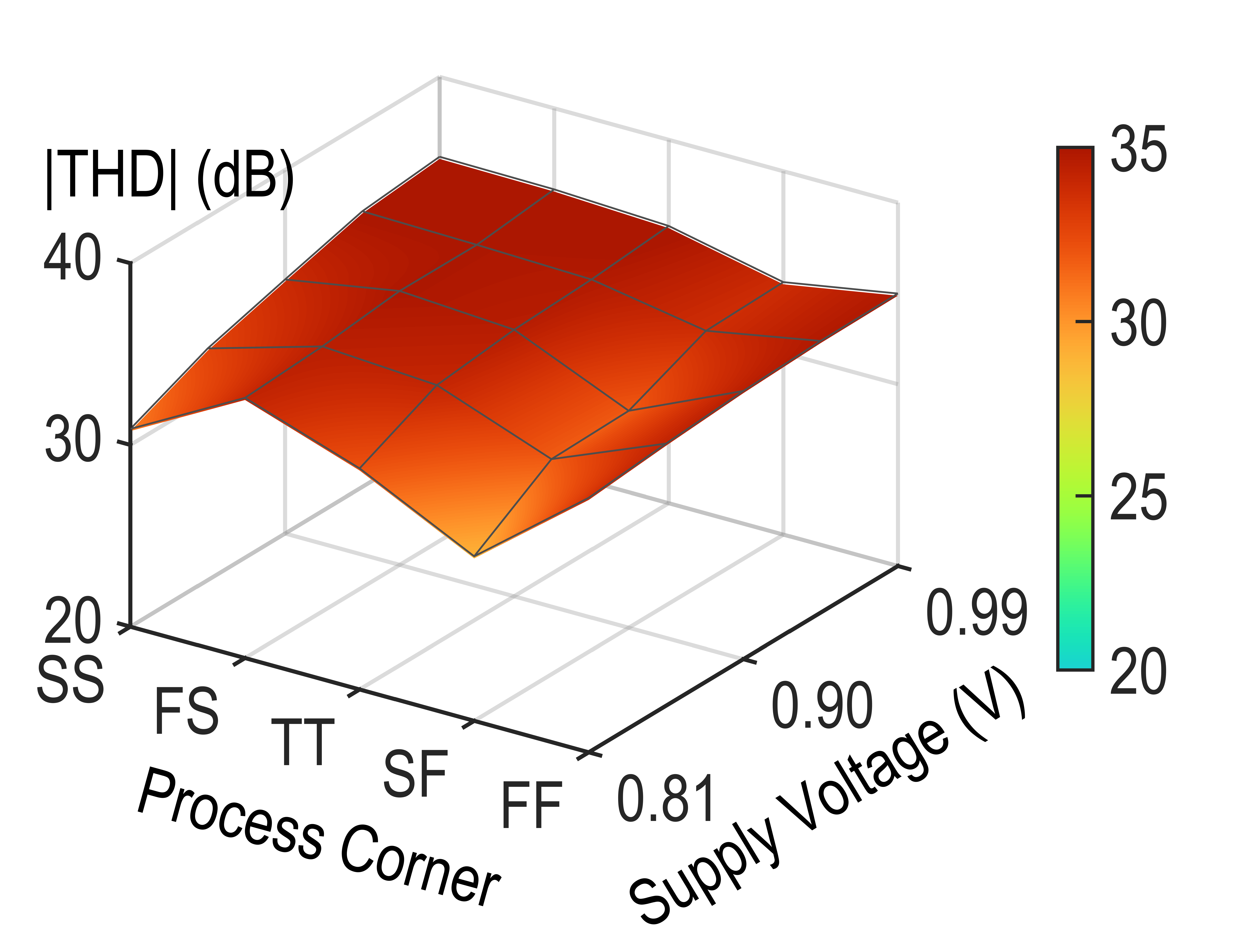}
        \vspace*{-1.8em}
        \caption{}
        \label{fig:pvt:f}
    \end{subfigure}
    \caption{Simulated midpoint offset, gain, and $|$THD$|$ versus process corner and supply voltage, (a)--(c) as programmed, with $s=26/31$ and the 0.58~V $\Vleak$ at every point, and (d)--(f) recentered, with the $\Vleak$ adjusted and the state reselected at each point.}
    \label{fig:pvt}
\end{figure*}

Process and supply also shift the transfer curve, and they change its gain. The VTC is simulated at 25 points, the SS, FS, TT, SF, and FF corners at each of five supply voltages spanning $\pm$10\,\% of 0.90~V. The midpoint is taken from the transfer curve for a dc input stepped from 0 to 0.80~V, the range that every supply covers. The midpoint offset is the shift of this midpoint from its value at the design point, the input-referred offset of Section~\ref{sec:vtc:fundamentals}. As programmed, every point holds $s=26/31$ and $\Vleak=$0.58~V of the design point. For recentering, $\Vleak$ of each point is adjusted in 5~mV steps to return the gain, and its state is reselected to return the midpoint, each to its value at the design point.

As programmed, the midpoint offset ranges from $-$61 to $+$22~mV and the gain from 0.18 to 2.29~ns/V, both set mainly by the corner [see Fig.~\ref{fig:pvt}(a-b)]. As a result, the THD degrades to $-$17.0~dB at SF and 0.81~V [see Fig.~\ref{fig:pvt}(c)]. After recentering, the midpoint offset falls to within 10~mV, about half the 17 to 20~mV by which one step of $s$ moves the midpoint, and the gain returns to within 4\,\% of 0.60~ns/V [see Fig.~\ref{fig:pvt}(d) and (e)]. The worst case remains SF at 0.81~V, where the THD improves to $-$28.9~dB, and the lowest supply now limits the THD at every corner [see Fig.~\ref{fig:pvt}(f)].

Adjusted alone, $\Vleak$ returns the gain but moves the midpoint by up to 41~mV, and reselected alone, the state returns the midpoint as closely as the two together but leaves the gain set by the corner. Together they recenter the transfer curve at every point, the state running from $s=24/31$ to $28/31$ and the bias from 0.49 to 0.66~V. At FF and FS, the state lies nearer HVT and the bias is lower than at SS and SF. At SS and SF, the slow leaker needs a higher bias to return the gain, and the slow FeFET at that bias needs a state closer to LVT to return the midpoint. The gain remains set by a bias voltage as in~\cite{ohhata2019}, while the offset trim, held outside the converting transistor in~\cite{ohhata2019,yi2021,qchen2022}, is here the programmed state of the FeFET that converts the input.

\section{Experimental Results}
\label{sec:results}

\begin{figure}[!t]
    \centering
    \includegraphics[width=0.6\columnwidth,trim={0.6cm 0.6cm 0.6cm 0.6cm},clip]{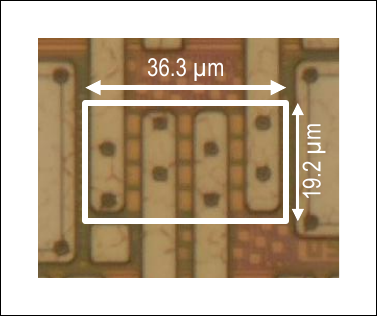}
    \caption{Die photograph of the proposed VTC.}
    \label{fig:chip}
\end{figure}

\begin{figure}[!t]
    \centering
    \includegraphics[width=\columnwidth,trim={0.6cm 0.6cm 0.6cm 0.6cm},clip]{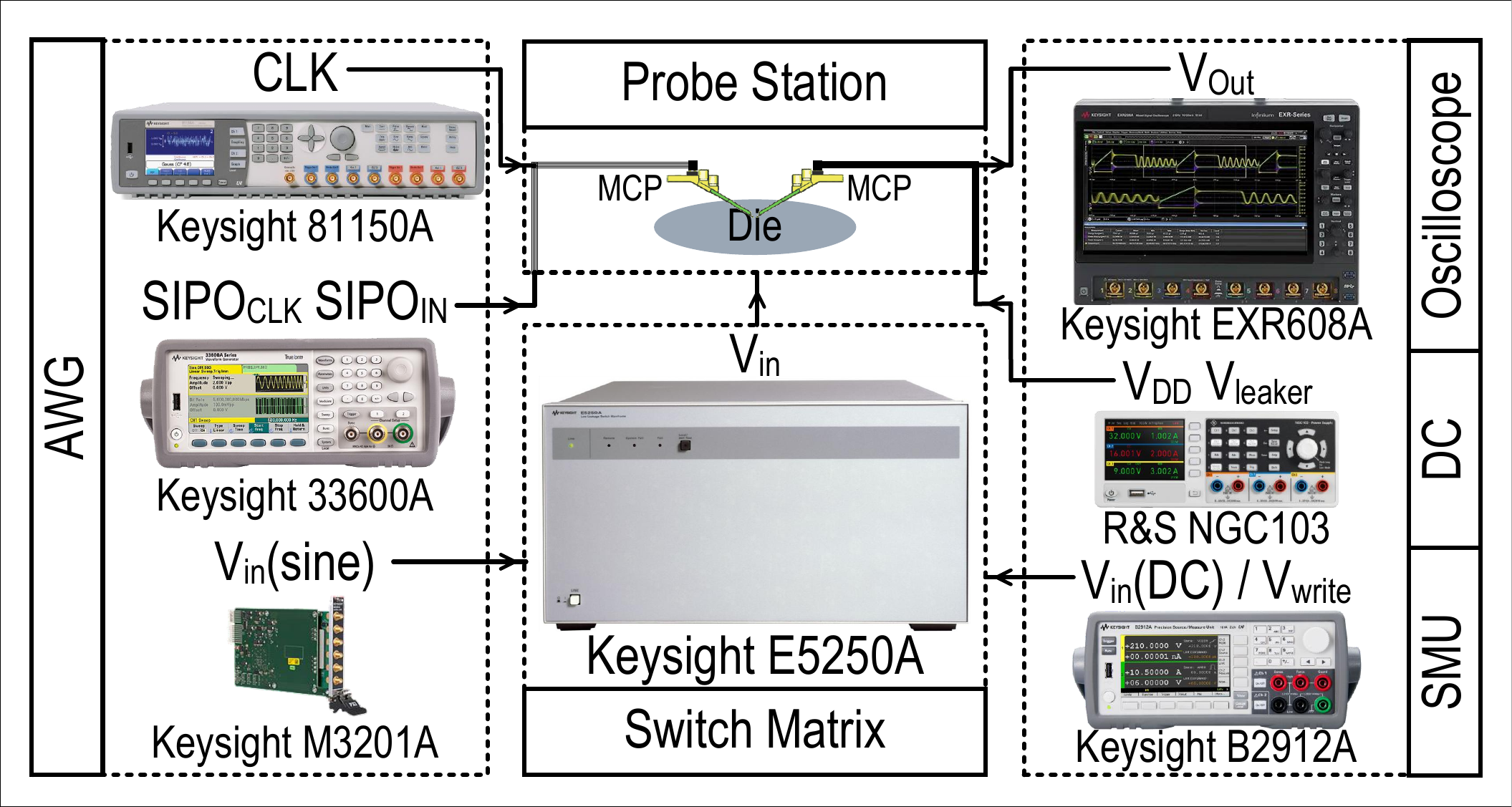}
    \caption{Measurement setup, with the die probed by two multi-contact probes (MCPs).
    }
    \label{fig:setup}
\end{figure}

The die area including the VTC, its I/O driver, and an on-chip serial-in parallel-out (SIPO) register that sets the bank code and the enable signals is 36.3~$\times$~19.2~$\mu$m$^2$, as shown in Fig.~\ref{fig:chip}. The VTC itself occupies an area of 6.07~$\mu$m$^2$. In the measurement setup of Fig.~\ref{fig:setup}, the clock, the SIPO bit stream, and the sine input are provided by arbitrary waveform generators (AWGs), while the supply voltage and the $\Vleak$ are maintained by a dc power supply. The FeFET gate is connected through a switch matrix to either a source-measure unit (SMU) for programming and dc input measurements or the sine AWG. The output is sampled by a 6~GHz oscilloscope through a dc-coupled 50~$\Omega$ input at 16~GSa/s, and the width of each low output pulse, $t_{PW}$ of \eqref{eq:width}, is measured. The VTC is operated at room temperature with a 10~MHz clock. The capacitor bank is set to code 0, corresponding to its smallest capacitance, unless the code is swept. The supply voltage, $\Vleak$, and input range are specified for each result.

\subsection{Programmed-State Transfer and Static Linearity}
\label{sec:static}

\begin{figure}[!t]
    \centering
    \includegraphics[width=0.8\columnwidth,trim={3cm 2.5cm 2.5cm 2.5cm},clip]{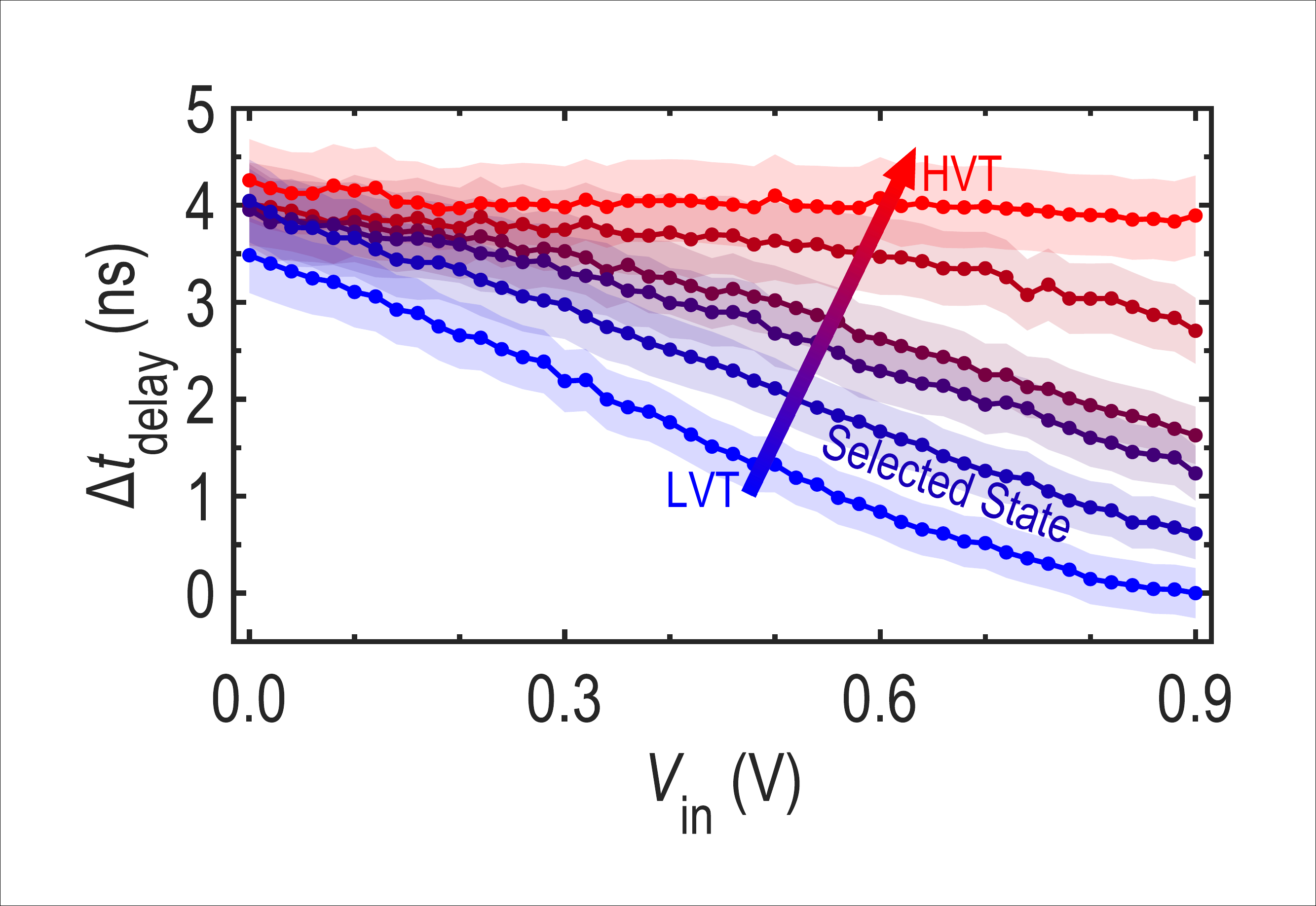}
    \caption{Measured $\Delta t_{delay}$ versus dc input for the six programmed states from LVT to HVT. The bands mark $\pm$1 standard deviation over repeated conversions.}
    \label{fig:family}
\end{figure}

The transfer curve of the VTC is measured with its FeFET in six programmed MLSs from LVT to HVT, at a 0.90~V supply and $\Vleak=0.44$~V, as shown in Fig.~\ref{fig:family}. In each state, the dc input is swept from 0 to 0.90~V in 20~mV steps. Each point is the mean $t_{PW}$ over repeated conversions, and the band marks their standard deviation, which ranges from 0.26 to 0.43~ns.

The delay is taken relative to that of LVT at a 0.90~V input, as $\Delta t_{delay}=t_{PW}(\mathrm{LVT},0.90~\mathrm{V})-t_{PW}$, removing the fixed offset of \eqref{eq:width}. The programmed state shifts the transfer curve and moves its midpoint from 0.40~V at LVT to 0.66~V at the intermediate state nearest HVT. The delay of HVT varies over the whole input range by less than 0.43~ns, which is about one standard deviation, so its midpoint lies above that range. The selected state is the one whose midpoint of 0.45~V lies nearest the center of the input range.

\begin{figure}[!t]
    \centering
    \includegraphics[width=0.8\columnwidth,trim={0.6cm 1.6cm 0.5cm 0.6cm},clip]{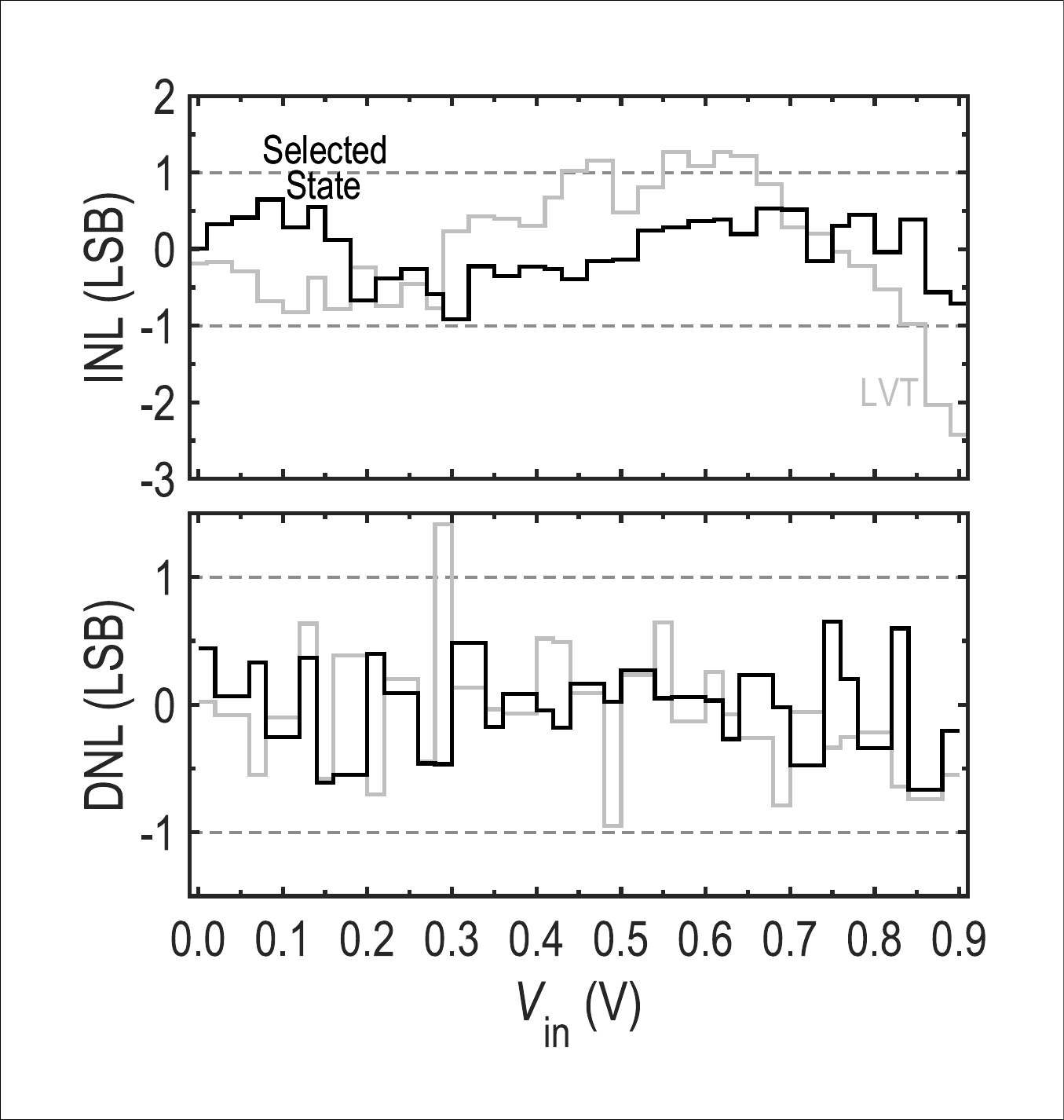}
    \caption{Measured INL and DNL of the selected state (black) and LVT (gray) for 5-bit codes over 0 to 0.90~V. The dashed lines mark $\pm$1~LSB.}
    \label{fig:linearity}
\end{figure}

The INL and differential nonlinearity (DNL) of the mean transfer curves of the selected state and LVT are shown in Fig.~\ref{fig:linearity} at 5-bit resolution over the input range. Each transfer curve is evaluated at the measured inputs nearest the code boundaries. The reference is the least-squares line of the mean $t_{PW}$ versus $V_{in}$ at these inputs. Its rise over one code is the LSB, 110~ps for the selected state and 117~ps for LVT. For the DNL, each step between adjacent inputs is divided by the rise of the line over the same inputs. Each point averages repeated conversions, so the noise of a single conversion adds a negligible random error to the INL and DNL. The uncertainty of a point is instead estimated from its scatter about the line through its two neighbors and is 23~ps for the selected state and 31~ps for LVT.

The selected state has an INL within $+$0.65/$-$0.92~LSB and a DNL within $+$0.65/$-$0.67~LSB, compared with $+$1.27/$-$2.42~LSB and $+$1.41/$-$0.95~LSB for LVT. The DNL of both states stays within a few times the uncertainty of a step, so only the INL separates the two states. The INL of LVT is worst at the highest inputs, reaching $-$2.42~LSB at 0.90~V, where the FeFET in its most conductive state enters the triode region earlier in each discharge phase, flattening the transfer curve. As in the simulations of Section~\ref{sec:sim}, where the centered midpoint gives the least distortion, the selected state has the smallest INL of the six, 0.92~LSB compared with 2.4~LSB or more for the other states.

\subsection{Transfer-Curve Placement and Single-Tone Response}
\label{sec:meas:sine}
\label{sec:meas:programmed}

\begin{figure}[!t]
    \centering
    \includegraphics[width=0.8\columnwidth,trim={2.5cm 2cm 1.0cm 1.0cm},clip]{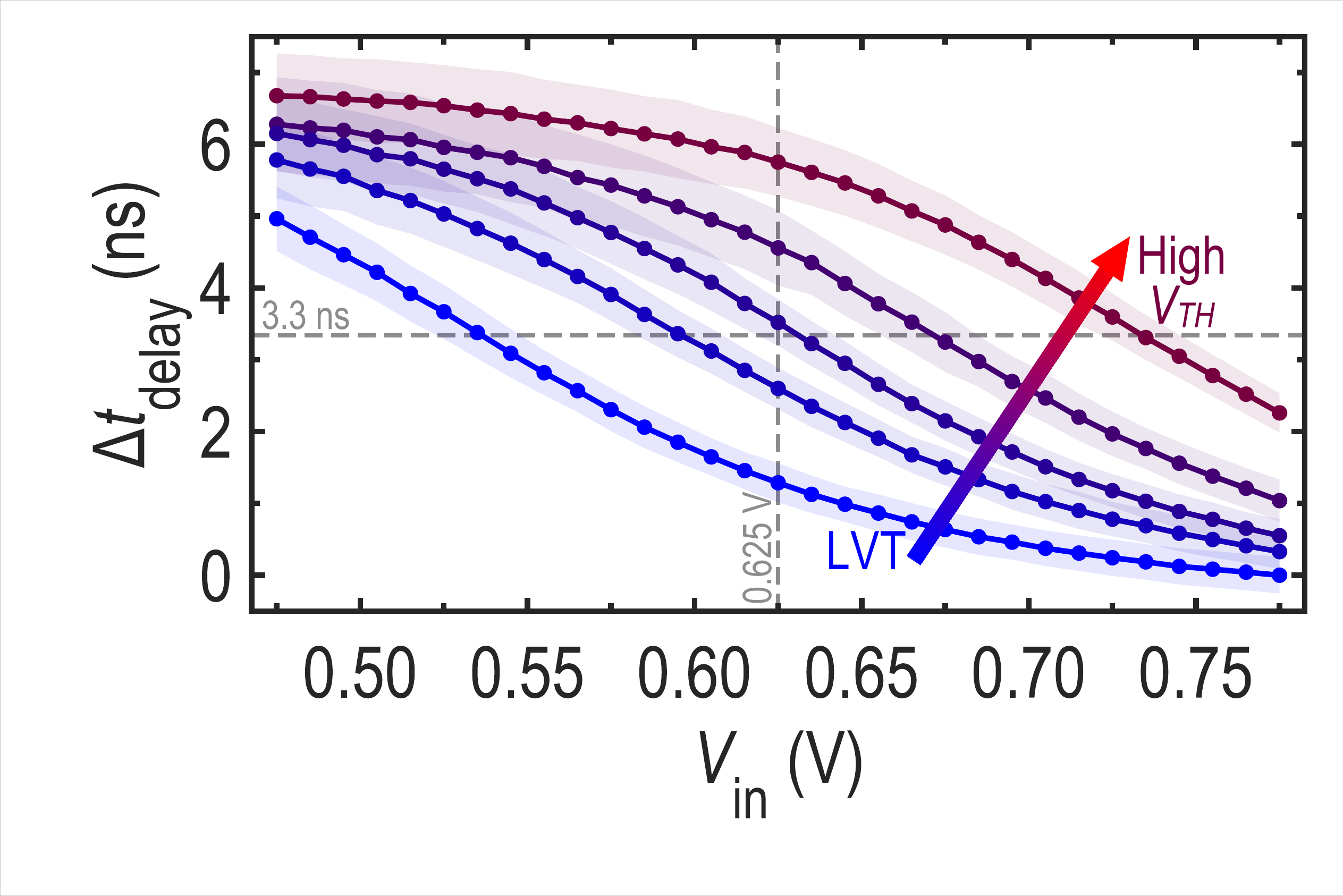}
    \caption{Measured $\Delta t_{delay}$ versus $V_{in}$ under the sine input, averaged over 10~mV intervals, for LVT and four intermediate states, with the delay taken relative to that of LVT at 0.78~V. The bands mark $\pm$1 standard deviation of $\Delta t_{delay}$ over the conversions in each interval. The dashed lines mark the 0.625~V center of the input sine and the 3.3~ns midpoint delay.}
    \label{fig:meas:sine}
\end{figure}

The VTC is measured under a sine input with its FeFET programmed to LVT and intermediate states, at a 0.90~V supply and $\Vleak=0.42$~V. The 98.8~kHz sine has a 0.31~V peak-to-peak swing centered on 0.625~V. In each state, the recorded widths $t_{PW}$ follow this sine in time, and a sine of amplitude $A$ is fitted to them. Since each width depends only on the input at its own conversion, the fitted sine has the same phase as the input. Together with the center and swing of the input sine, this phase gives the input voltage $V_{in}$ of every conversion. The widths are then sorted by $V_{in}$ into 10~mV intervals, and the mean of each interval forms one point of the transfer curve. These curves are shown in Fig.~\ref{fig:meas:sine} for LVT and four intermediate states, with $\Delta t_{delay}$ taken relative to the delay of LVT at the 0.78~V upper peak of the sine.

\begin{figure}[!t]
    \centering
    \includegraphics[width=0.8\columnwidth,trim={2.8cm 3cm 1.0cm 1.0cm},clip]{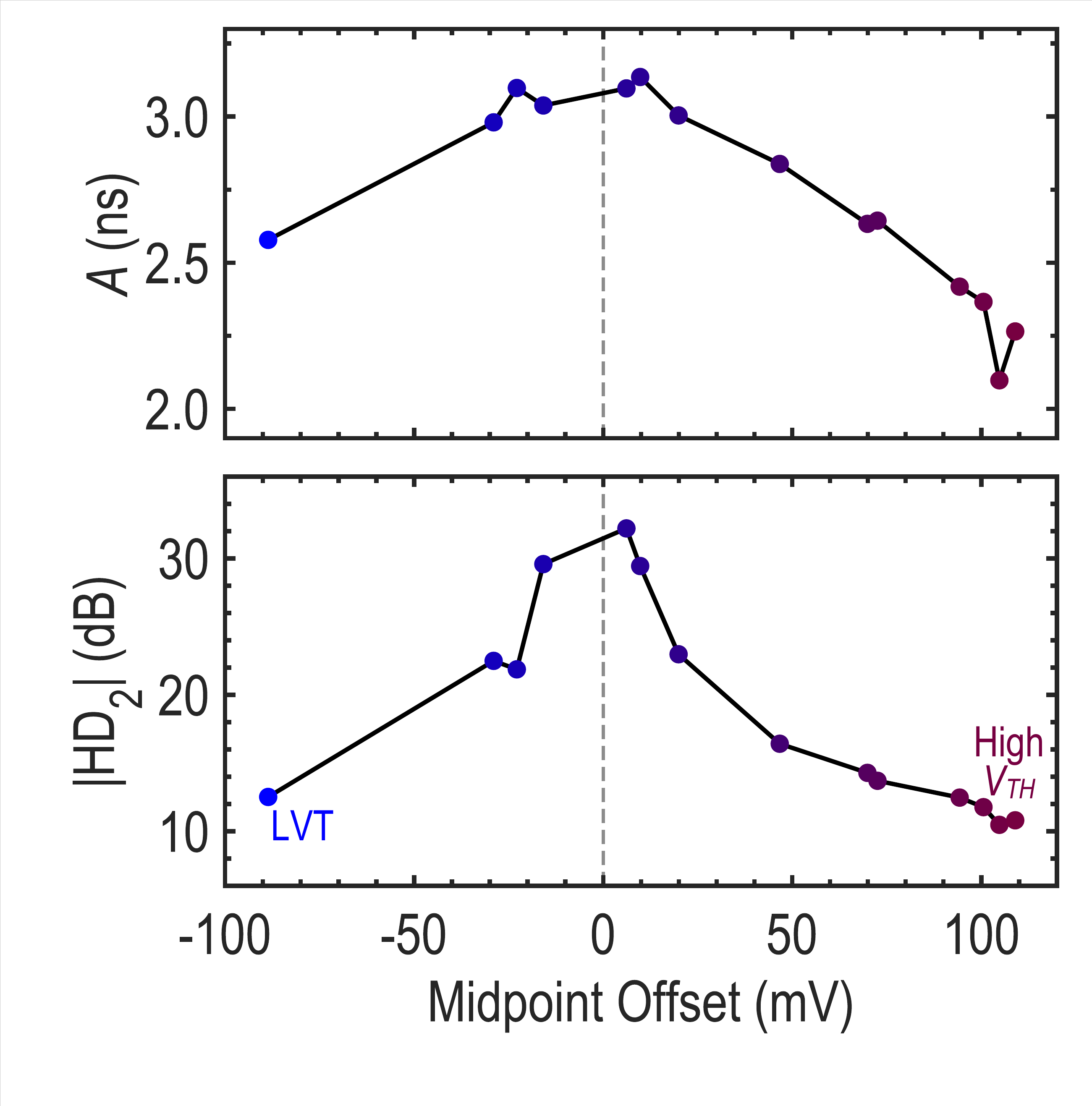}
    \caption{Measured amplitude $A$ and second harmonic ($|\mathrm{HD}_2|$) of $t_{PW}$ versus the midpoint offset for LVT and 13 intermediate states. The dashed line marks zero offset.}
    \label{fig:meas:programmed}
\end{figure}

Each intermediate state shifts every point of the transfer curve toward higher inputs relative to LVT by nearly the same amount, so the programmed state acts as an input-referred offset trim. The midpoint delay of 3.3~ns lies halfway between the flat ends of the curves at 0 and 6.7~ns. The position of each curve on the input sine is its midpoint offset, the input at which $\Delta t_{delay}$ crosses 3.3~ns measured from the 0.625~V center of the sine. A curve through the intersection of the two dashed lines in Fig.~\ref{fig:meas:sine} has zero offset and is centered on the sine. LVT has an offset of $-$89~mV, while the offsets of the intermediate states lie on both sides of zero.

The amplitude $A$ and the second harmonic of each state are shown against the midpoint offset of that state in Fig.~\ref{fig:meas:programmed}. Within $\pm$30~mV of zero offset, $A$ remains within 3\,\% of 3.06~ns. At larger offsets, the sine reaches a flat end of the curve, and $A$ falls to as low as 2.10~ns. The second harmonic is weakest in the intermediate state with an offset of $+$6~mV, where it is $-$32.2~dB against $-$12.5~dB in LVT, and it grows as the offset moves away from zero in either direction.

A peak of the input sine is compressed in the output when that peak approaches a flat end of the curve, where the delay responds less to the input. The second harmonic follows from how unequally the two peaks are compressed. With a negative offset, the upper peak is compressed more than the lower one, and a positive offset compresses the lower peak more. Away from zero offset, the second harmonic dominates the THD, which reaches $-$12.4~dB in LVT. Near zero offset, where the state centers the curve on the sine, both peaks are compressed about equally, and the second harmonic nearly cancels, as in the simulations. The equal compression of both peaks still distorts the output through the third harmonic, which dominates within $\pm$30~mV of zero offset, where the THD stays between $-$18.4 and $-$19.9~dB. The centered state is the one with the $+$6~mV offset. Its THD is $-$19.6~dB and its SFDR is 19.9~dB.

In the centered state, the resolution follows from \eqref{eq:enob} to \eqref{eq:snr}, with $\Delta t_{PW}=2A=6.19$~ns and $\sigma_{PW}=0.38$~ns. Under the sine input, $\sigma_{PW}$ is estimated as the rms residual of $t_{PW}$ after the fitted sine and its harmonics are subtracted. The SNR is then 15.2~dB, and with the distortion added, the SNDR is 13.9~dB and the ENOB 2.0~bit. The states within $\pm$30~mV of zero offset give an SNDR of 11.7 to 13.9~dB, the highest in the centered state. In each of these states, the noise power exceeds the distortion power. In LVT, the distortion power instead exceeds the noise power, and the SNDR is 10.6~dB, an ENOB of 1.5~bit. Centering the transfer curve with the programmed state moves the limit on the resolution from the distortion in LVT to the noise near zero offset.

\subsection{Transfer-Curve Gain by Bank Code and Leaker Bias}
\label{sec:meas:controls}

\begin{figure}[!t]
    \centering
    \begin{subfigure}{\columnwidth}
        \centering
        \includegraphics[width=0.88\columnwidth, trim={1.5cm 1.5cm 1.5cm 1.5cm}, clip]{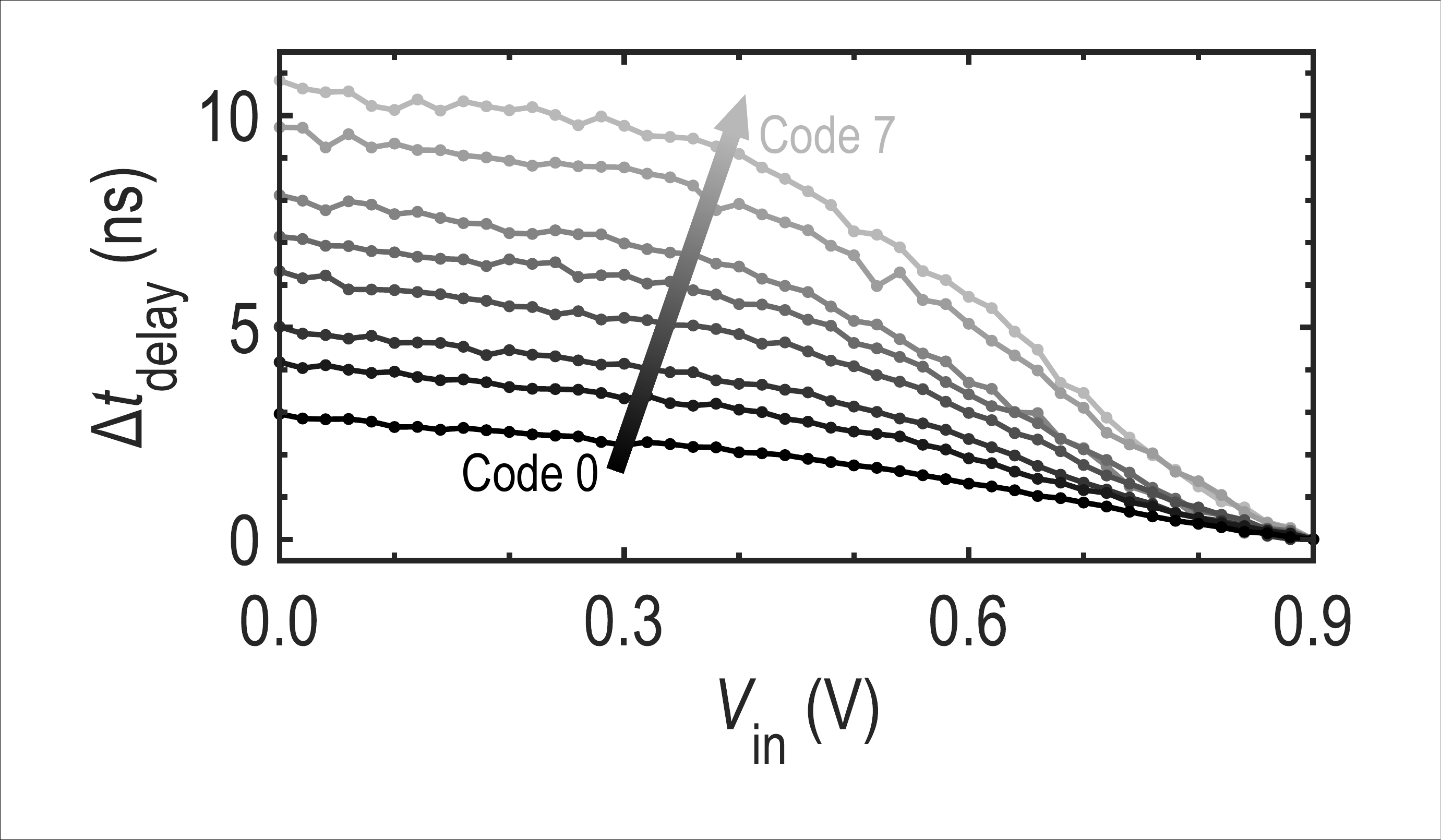}
        \vspace{-0.6em}
        \caption{}
        \label{fig:meas:controls:a}
    \end{subfigure} \\
    \vspace{-0.1em}
    \begin{subfigure}{\columnwidth}
        \centering
        \includegraphics[width=0.85\columnwidth, trim={1cm 1cm 1cm 1cm}, clip]{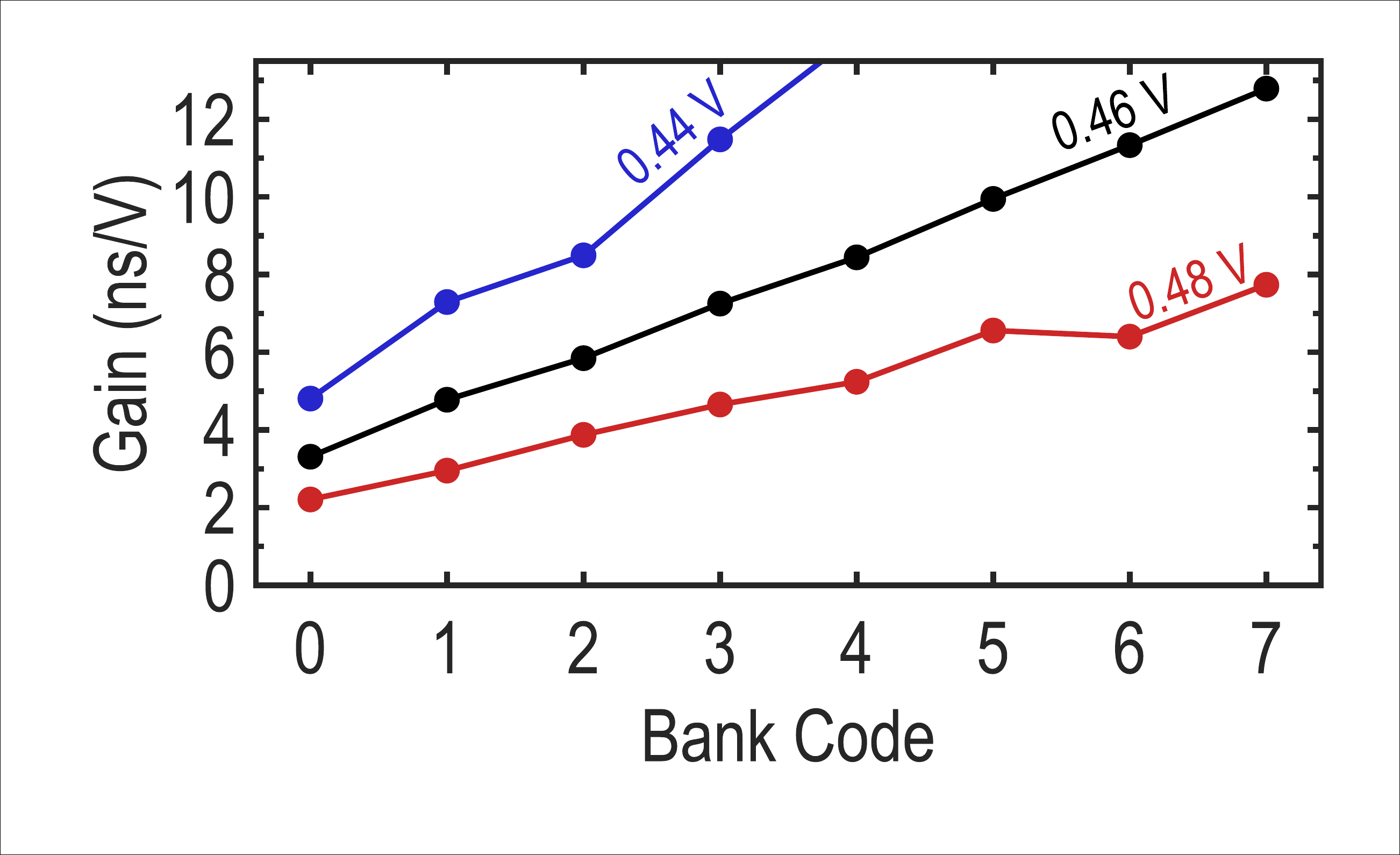}
        \vspace*{-1em}
        \caption{}
        \label{fig:meas:controls:b}
    \end{subfigure}
    \caption{Measured static transfer curves with the FeFET in LVT at a 0.90~V supply. (a) $\Delta t_{delay}$ versus $V_{in}$ for bank codes 0 (black) to 7 (light gray) at $\Vleak=0.46$~V, each relative to its own delay at 0.90~V. (b) Least-squares gain over 0 to 0.90~V versus bank code at $\Vleak$ of 0.44, 0.46, and 0.48~V.}
    \label{fig:meas:controls}
\end{figure}

The effects of the capacitor-bank code and $\Vleak$ on the static transfer characteristic are measured with the FeFET in LVT at a 0.90~V supply. At a $\Vleak$ of 0.46~V, the midpoint remains between 0.57 and 0.62~V across bank codes 0 to 7 [see Fig.~\ref{fig:meas:controls}(a)]. The least-squares gain over the full 0--0.90~V input range is increased from 3.3~ns/V at code 0 to 12.8~ns/V at code 7 [see Fig.~\ref{fig:meas:controls}(b)]. When $\Vleak$ is increased from 0.44 to 0.48~V, the gain falls to approximately one-third to one-half of its value at 0.44~V. The bank code and $\Vleak$ thus provide gain adjustment, while the programmed state controls transfer-curve placement at nearly constant gain.

The dependence of SNR on $\Vleak$ is also evaluated under a sinusoidal input. With the FeFET programmed to LVT before each capture, $\Vleak$ is swept from 0.42 to 0.56~V in 20~mV steps under a 0.30~V peak-to-peak sine centered on 0.625~V. Across this sweep, the pulse-width swing falls by a factor of 14.7, whereas $\sigma_{PW}$ falls by a factor of 4.8. Consequently, the SNR decreases with each increase in bias. A $\Vleak$ of 0.42~V, which gives the highest SNR in this sweep, is therefore used for the single-tone measurements of Section~\ref{sec:meas:sine}.

\section{Discussion and Comparison With Prior Art}
\label{sec:comparison}

\begin{table*}[!t]
    \centering
    \footnotesize
    \renewcommand{\arraystretch}{1.2}
    \setlength{\tabcolsep}{3pt}
    \caption{\textsc{Comparison With VTCs and VTC-Based ADCs That Carry Offset and Gain Trims}}
    \label{tab:compare_vtc}
    \begin{tabular}{>{\raggedright\arraybackslash}p{2.6cm}C{1.9cm}C{1.9cm}C{1.9cm}C{1.9cm}C{1.9cm}C{1.75cm}C{1.75cm}}
        \textbf{Reference} & JSSC'19~\cite{ohhata2019} & JSSC'21~\cite{yi2021} & Access'22~\cite{qchen2022} & JSSC'23~\cite{zhao2023} & ISSCC'26~\cite{wang2026} & \multicolumn{2}{c}{\textbf{This Work}} \\
         & & & & & & \textbf{Meas.}$^{\mathrm{a}}$ & \textbf{Sim.}$^{\mathrm{b}}$ \\ \hline\hline
        Offset trim & Bias-voltage difference & 4-bit load capacitance & 4-bit delay buffer & 4-bit capacitor bank & TDC phase-interpolator offsets & \multicolumn{2}{C{\dimexpr3.5cm+2\tabcolsep\relax}}{\textbf{Converting FeFET state}} \\ \hline
        Offset trimmed as & Not stated & Time offset & Time offset & Not stated & Time offset & \multicolumn{2}{C{\dimexpr3.5cm+2\tabcolsep\relax}}{\textbf{Input-referred offset}} \\ \hline
        Gain set by & Bias voltages & Time-amplifier gain & Bias voltage, 3-bit charging path & 4-bit capacitor bank & Background bias loop & \multicolumn{2}{C{\dimexpr3.5cm+2\tabcolsep\relax}}{\textbf{Leaker bias, 3-bit capacitor bank}} \\ \hline\hline
        Architecture & Two-step ADC & Time-based ADC & Constant-slope VTC & Interleaved SAR ADC & Pipelined-SAR ADC & \multicolumn{2}{C{\dimexpr3.5cm+2\tabcolsep\relax}}{\textbf{Variable-slope VTC}} \\ \hline
        Technology (nm) & 65 & 65 & 28 & 28 & 28 & \multicolumn{2}{c}{\textbf{28}} \\ \hline
        Supply (V) & 1 & 1 & 0.6 & 0.9 & 1 & \multicolumn{2}{c}{\textbf{0.90}} \\ \hline
        Input range (V$_{\mathrm{pp}}$) & 0.88 diff. & 0.8 diff. & 0.6 & 0.8 diff. & 1.8 diff. & {\textbf{0.31}} & {\textbf{0.40}} \\ \hline
        Rate (MS/s) & 1000 & 4000 & 4000 & 2800 & 400 & {\textbf{10}} & {\textbf{500}} \\ \hline
        Power$^{\mathrm{c}}$ & 2.3~mW & 11.3~mW & 2.1~mW & 18.05~mW & 4.59~mW & {\textbf{--}} & {\textbf{2.0~$\mu$W}} \\ \hline
        Area ($\mu$m$^2$)$^{\mathrm{c}}$ & 7000 & 11\,000 & 1200 & 9700 & 16\,000 & \multicolumn{2}{c}{\textbf{6.07}} \\ \hline\hline
        \multicolumn{8}{@{}p{\dimexpr15.6cm+14\tabcolsep\relax}@{}}{$^{\mathrm{a}}$Measured at room temperature, $\Vleak=0.42$~V, 98.8~kHz sine, centered state.}\\
        \multicolumn{8}{@{}p{\dimexpr15.6cm+14\tabcolsep\relax}@{}}{$^{\mathrm{b}}$Post-layout simulation at TT, 70~$^\circ$C, $\Vleak=0.58$~V, 19.53~MHz sine, ideal clock and input sources.}\\
        \multicolumn{8}{@{}p{\dimexpr15.6cm+14\tabcolsep\relax}@{}}{$^{\mathrm{c}}$Of the block in the Architecture row, for~\cite{qchen2022} with clock generation and for its area also with output drivers.}\\
    \end{tabular}
\end{table*}
Measurement and simulation show consistent effects of the programmed MLS on the transfer characteristic. In the measured curves, each intermediate state produces a nearly uniform shift toward higher input voltages relative to LVT. Within $\pm$30~mV of zero offset, the fitted pulse-width amplitude remains nearly constant. By centering the transfer characteristic, the programmed state reduces the second harmonic in simulation and in measurement (see Sections~\ref{sec:sim:state} and~\ref{sec:meas:sine}). The measured second harmonic increases as the offset departs from zero in either direction. At $\Vleak=0.44$~V, the state whose midpoint lies nearest the center of the input range has the smallest measured static INL (see Section~\ref{sec:static}).

Simulation and measurement use different operating points, and performance at the design point is consequently evaluated only through simulation. For the centered state, a gain of 20~ns/V, defined as the ratio of output width swing to input voltage swing, is measured with a 0.31~V peak-to-peak sine input, $\Vleak=0.42$~V, and room-temperature operation. The optimum state has a simulated gain of 0.60~ns/V at $\Vleak=0.58$~V and 70~$^\circ$C. A higher $\Vleak$ reduces this simulated gain and, similarly, the measured least-squares gain in LVT over the full input range. The simulated gain varies by less than 3\,\% between 100 and 700~MS/s. In measurement, the 6.19~ns output width swing covers nearly the entire 6.7~ns span between the flat ends of the transfer curve. Both sine peaks are therefore compressed, and the third harmonic dominates the measured THD of $-$19.6~dB. In simulation, THD remains below $-$30~dB for input swings up to 0.50~V peak-to-peak, with the 0.40~V input swing confined away from the flat ends.

Near zero offset, noise rather than distortion limits the conversion resolution in simulation and measurement. The measured $\sigma_{PW}$ of 0.38~ns, the rms residual after subtraction of the fitted sine and its harmonics, is 84 times the 4.5~ps simulated at a constant input with ideal sources. The mean conversion delay, however, is 28 times longer in the centered state than at the design point. Each $\sigma_{PW}$ is therefore divided by this delay, evaluated as the discharge phase minus the mean width. This relative noise is 1.3\,\% at the design point and 3.9\,\% in the centered state, accounting for the remaining factor of three. Across the $\Vleak$ sweep of Section~\ref{sec:meas:controls}, $\sigma_{PW}$ and the delay decrease 4.8-fold while the gain decreases 14.7-fold, and the relative noise remains between 3.9 and 4.7\,\%. Noise proportional to the delay, rather than to the gain, points to a relative fluctuation of the discharge current. The larger measured relative noise indicates a source absent from the simulation, such as noise on $\Vleak$ from the dc power supply or low-frequency leaker noise.

The offset trim of this VTC relies on the stability of the programmed state. On the same platform, FeFETs with a 90~nm gate width and length show negligible drift of their pulsed read current in several MLSs over 10$^5$~s at room temperature, and that current stays consistent up to about 10$^3$ full-swing cycles~\cite{mattar2026}. Four levels written to FeFETs with the 450~nm gate width and length used here stay separated after 24~h at 85~$^\circ$C~\cite{muller2023percolation}, and a binary 28-nm FeFET array keeps its pattern after 72~h at 105~$^\circ$C and reaches 10$^5$ cycles~\cite{trentzsch2016}.

The proposed VTC is compared with trimmed VTCs and VTC-based ADCs in Table~\ref{tab:compare_vtc}, with emphasis on their offset-trimming mechanisms. In~\cite{yi2021}, a digitally coded load capacitance adjusts the time offset in steps of 0.34~LSB, whereas a digitally coded delay buffer provides 500~fs steps in~\cite{qchen2022}. Both mechanisms are implemented outside the converting transistor and add a uniform delay to the output without shifting the transfer-curve midpoint, as discussed in Section~\ref{sec:vtc:fundamentals}. Consequently, an additive time offset cannot compensate for the simulated THD degradation of 3.5 to 6.1~dB caused in the proposed VTC by a 50~mV midpoint displacement. In the proposed circuit, the programmed FeFET state trims the input-referred offset. At a $\Vleak$ of 0.44~V, the measured static midpoint shifts from 0.40~V in LVT to 0.66~V in the intermediate state nearest HVT. Across simulated process corners and supply variations, recentering reduces the midpoint offset from an initial range of $-$61 to $+$22~mV to within 10~mV. This residual is approximately half the 17--20~mV midpoint shift produced by one step of $s$.

In the compared ADCs, gain is controlled through bias voltages adjusted manually for each measurement~\cite{ohhata2019}, a time-amplifier gain configured manually once~\cite{yi2021}, a capacitor bank configured during foreground calibration~\cite{zhao2023}, or a background bias-control loop~\cite{wang2026}. In the proposed VTC, $\Vleak$ and the capacitor-bank code adjust the gain. With the FeFET in LVT and $\Vleak$ fixed at 0.46~V, varying the bank code increases the measured least-squares gain from 3.3 to 12.8~ns/V. The fabricated VTC occupies an area of 6.07~$\mu$m$^2$. For comparison, the work in~\cite{qchen2022} reports a 21~$\mu$m~$\times$~30~$\mu$m core within the 1200~$\mu$m$^2$ active area of its pseudo-differential constant-slope VTC, in which the input is sampled through bootstrapped switches. Compared works do not report standalone VTC area.

\begin{table}[!t]
    \centering
    \footnotesize
    \renewcommand{\arraystretch}{1.2}
    \setlength{\tabcolsep}{1.5pt}
    \caption{\textsc{Comparison With Circuits Set by a Stored Device State}}
    \label{tab:compare_nv}
    \begin{tabular}{>{\raggedright\arraybackslash}p{1.83cm}C{1.4cm}C{1.5cm}C{0.9cm}C{0.9cm}C{1.7cm}}
        \textbf{Reference} & Stored as & Held in & Input & Output & One state sets \\ \hline\hline
        TCAS-I'08~\cite{wong2008comp} & FG charge & Comparator input pair & Analog & Binary & One input offset \\ \hline
        EDL'21~\cite{nouri2021ring} & Trapped charge & Ring transistors & -- & Freq. & One freq. \\ \hline
        SSC-L'26~\cite{mema2026} & Binary FeFET state & RO tail FeFETs & -- & Freq. & All stage delays \\ \hline
        TCAS-I'09~\cite{wong2009quant} & FG charge & Comparator inputs & Analog & Code & One reference level \\ \hline
        NC'25~\cite{yang2025}\textsuperscript{a} & ReRAM conductance & Ramp-generator ReRAMs & Analog & Code & One ramp step \\ \hline
        NC'25~\cite{hong2025adaptiveadc}\textsuperscript{b} & ReRAM conductance & Quantizer ReRAMs & Analog & Code & One boundary \\ \hline
        TCAS-I'26~\cite{mattar2026} & FeFET MLS & Inverter-tail FeFETs & Binary & Delay & One delay \\ \hline
        \textbf{This Work} & \textbf{FeFET MLS} & \textbf{Inverter-tail FeFET} & \textbf{Analog} & \textbf{Delay} & \textbf{Transfer-curve offset} \\ \hline\hline
        \multicolumn{6}{@{}p{\dimexpr7.8cm+10\tabcolsep\relax}@{}}{\textsuperscript{a}Array measured, integrator and comparator in software.}\\
        \multicolumn{6}{@{}p{\dimexpr7.8cm+10\tabcolsep\relax}@{}}{\textsuperscript{b}Simulated converter.}\\
    \end{tabular}
    \vspace*{-1.5em}
\end{table}

Circuits configured through stored device states are compared in Table~\ref{tab:compare_nv}, with emphasis on the quantity controlled by each state. In the comparator of~\cite{wong2008comp}, charge stored on the floating-gate input pair, to which the input is capacitively coupled, sets an arbitrary input offset. As in the proposed VTC, the offset is therefore stored in the input device, although the comparator produces a binary decision rather than a time interval. In the ring oscillator (RO) of~\cite{nouri2021ring}, charge trapped in the high-$k$ gate dielectric of its transistors tunes the frequency. In the current-starved ring oscillator of~\cite{mema2026}, the programmed states of a p-FeFET and an n-FeFET, whose gates are held at fixed binary biases, control the stage delay. In the flash quantizer of~\cite{wong2009quant}, an array of floating-gate comparators stores 63 reference levels. Similarly, separate ReRAM devices encode individual ramp steps~\cite{yang2025} or quantization boundaries~\cite{hong2025adaptiveadc}. In these quantizers and converters, multiple devices encode the transfer characteristic. In the proposed VTC, the programmed state of a single FeFET positions a circuit-defined transfer characteristic.

The time-domain computing macro of~\cite{mattar2026} differs from this VTC in function and evaluation. In operation, its inputs are binary and each state trims one delay. The work in~\cite{mattar2026} measures the delay against a gate voltage swept from $-$0.2 to 1.8~V across MLSs but does not report gain, linearity, or distortion of that delay against the gate voltage. As in this VTC, its delay element is a current-starved inverter loaded by a capacitor bank, with FeFETs and a parallel NMOS leaker in its tail. Here, an analog input within the 0.90~V supply drives the gate of a single FeFET, and the INL and the harmonics of the transfer curve are measured in LVT and intermediate states. None of the compared circuits places the whole transfer curve of an analog-to-time converter on its input range by the programmed state of the device that converts. The measurements establish the offset trim held in the FeFET that converts the input, and the simulations add its recentering across process and supply.

\section{Conclusion}
\label{sec:conclusion}

A 6.07-$\mu$m$^2$ FeFET-based VTC with nonvolatile input-referred offset trim has been demonstrated in 28-nm CMOS. The programmed MLS positions the transfer characteristic over the input range, while the capacitor bank and leaker bias provide gain adjustment. Centering reduces the measured second harmonic from $-$12.5 to $-$32.2~dB and the maximum absolute static INL from 2.42 to 0.92~LSB when expressed in 5-bit LSB units. In post-layout simulations across process corners and supply variation, joint state and bias adjustment restores the midpoint to within 10~mV and the gain to within 4\% of nominal. Conversion resolution remains primarily limited by noise.

The proposed circuit could serve reconfigurable VTC--TDC front ends, where offset trim aligns the input range with the region of lowest distortion and gain adjustment matches the output time swing to the TDC range. Its stored state could also support per-element calibration of analog voltage-to-delay encoders in time-domain computing and persistent delay settings in programmable timing circuits. Further characterization of programming repeatability, retained-offset stability, and noise would help establish its suitability for these applications.

\section*{Acknowledgments}
The authors thank Prof. E. Yalon, Y. Keller, and I. Lipschutz for their support with the experimental measurements. 
\ifCLASSOPTIONcaptionsoff
  \newpage
\fi

\bibliographystyle{ieeetr}
\bibliography{references}

\end{document}